\documentclass[aps,prb,amsmath,reprint,amssymb,superscriptaddress]{revtex4-2}
\usepackage{amsmath}
\usepackage{graphicx}
\usepackage{dcolumn}
\usepackage{bm}
\usepackage{hyperref}

\usepackage{comment}
\usepackage{xcolor}

\begin{document}


\title{Magnetic-fluctuation-driven suppression of spin-orbit hybridization \\in the surface ferromagnet GdAg$_2$/Ag(111)}

\author{Ryo~Noguchi}
\email{rnoguchi@ntu.edu.tw}
\affiliation{Center for Correlated Electron Systems, Institute for Basic Science, Seoul, 08826 South Korea}
\affiliation{Center for Condensed Matter Sciences, National Taiwan University, Taipei 10617, Taiwan}
\affiliation{Center of Atomic Initiatives for New Materials, National Taiwan University, Taipei 10617, Taiwan}

\author{Jongkeun Jung}
\affiliation{Department of Physics and Astronomy, Seoul National University, Seoul, 08826 South Korea}

\author{Younsik Kim}
\affiliation{Department of Physics and Astronomy, Seoul National University, Seoul, 08826 South Korea}

\author{Sungsoo Hahn}
\affiliation{Department of Physics and Astronomy, Seoul National University, Seoul, 08826 South Korea}

\author{Changyoung~Kim}
\email{changyoung@snu.ac.kr}
\affiliation{Department of Physics and Astronomy, Seoul National University, Seoul, 08826 South Korea}

\date{\today}

\begin{abstract}
Magnetic materials hosting topological band structures have attracted intense interest due to the interplay between magnetism and spin-orbit coupling (SOC). Here, using temperature- and polarization-dependent angle-resolved photoemission spectroscopy, we investigate the surface ferromagnet GdAg$_2$/Ag(111), a two-dimensional system with Weyl-nodal-line-like band crossings. We find that spin fluctuations preserve the nodal-line-derived band crossings even above the Curie temperature, while SOC-induced hybridization becomes well-defined only at low temperatures, as evidenced by spectral-weight redistribution. The suppression of the hybridization at high temperature is attributed to spin decoherence and band-dependent scattering, captured by an effective non-Hermitian framework. Our results establish magnetic fluctuations as a control knob for SOC-induced hybridization and associated Berry curvature, and highlight magnetic systems as a platform for exploring non-Hermitian band physics.
\end{abstract}

\maketitle


The interplay between topological electronic structures and magnetism has emerged as a central theme in condensed matter physics, as it gives rise to a variety of unconventional transport phenomena \cite{Tokura2019,Chang2023}. Notable examples include the quantum anomalous Hall effect in magnetic topological insulators and large anomalous Hall responses in Weyl semimetals \cite{Chang2013a,Checkelsky2014,Otrokov2019,Gong2018a,Deng2020,Kuroda2017b,Liu2019d,Morali2019,Belopolski2019}. In particular, Weyl-type electronic states have attracted considerable attention as a key mechanism for generating large Berry curvature and associated transport responses \cite{Kuroda2017b,Liu2019d,Morali2019,Belopolski2019}. In magnetic systems, where exchange splitting lifts spin degeneracy, spin-orbit coupling (SOC) can strongly modify the electronic structure near band crossings, thereby controlling the Berry curvature distribution and transport properties \cite{Xu2018aa,Liu2018aa,Liu2022}. Understanding and controlling SOC-induced band hybridization in magnetic materials is therefore crucial for designing Berry curvature-based transport phenomena.

\begin{figure}[tb]
\begin{center}
\includegraphics[width=1\columnwidth]{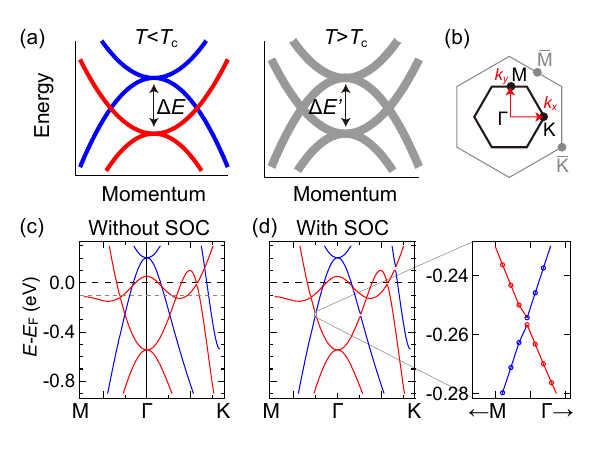}
\caption[]{(a) Schematics of spin-mixing-type band splitting in a ferromagnet with localized spin moments. The exchange splitting $\Delta E$ remains finite above $T_\mathrm{C}~(\Delta E^\prime)$, enabling band crossings to survive even in the paramagnetic state.  (b) Surface Brillouin zones of GdAg$_2$ (black) and the Ag(111) substrate (gray). (c) Theoretical band structures for a GdAg$_2$ monolayer obtained by density-functional theory calculations without SOC. The red and blue curves denote spin-up and spin-down bands, respectively. The gray dotted line indicates the approximate position of the experimental Fermi level. (d) Theoretical band structures calculated with SOC (left panel). The right panel shows the magnified dispersion around a band crossing point.}
\label{fig1}
\end{center}
\end{figure}

However, the temperature evolution of Weyl-type electronic states and SOC-induced band hybridizations exhibits system-dependent behavior, and a unified understanding remains elusive. One of the key factors is spin fluctuations \cite{Belopolski2021PRL,DFLiu2021,Rossi2021,Ma2019,Milo2024}. While spin splitting disappears above the Curie temperature ($T_{\rm C}$) in itinerant ferromagnets, it can persist even above $T_{\rm C}$ in systems with localized magnetic moments due to spin fluctuations [Fig. \ref{fig1}(a)], often referred to as a spin-mixing mechanism not captured by conventional density functional theory (DFT) \cite{Li1995,Weschke1996,Donath1996,Getzlaff1998,Maiti2002,Fedorov2002,Qin2022,Wu2023,Zhong2023}. Recent studies have shown that spin fluctuations can stabilize Weyl-type electronic states in Eu-based compounds above $T_{\rm C}$, underscoring the importance of spin fluctuations in shaping topological band structures \cite{Ma2019,Milo2024}. In contrast, the effect of spin fluctuations on SOC-induced hybridizations has not been systematically investigated, whereas spin-orientation dependence of SOC gaps has been resolved in a few materials \cite{Cheng2023,Lou2024}. 
Experimentally, resolving such small energy scales remains challenging. In angle-resolved photoemission spectroscopy (ARPES), SOC-induced gaps are often difficult to detect due to $k_z$ broadening \cite{Dama2004}, and their interpretation is further complicated by possible discrepancies between surface and bulk magnetic order \cite{Sanchez-Barriga2016,Shikin2020,Garnica2022}.

A promising platform to overcome these difficulties is provided by two-dimensional electronic systems realized in surface alloys \cite{Chuang2016,Feng2017c,Feng2019a,Unzelmann2020a,Liu2020,Lu2021,Bauernfeind2021,Cameau2024,Matetskiy2025}. A representative example is the surface alloy GdAg$_2$, which exhibits ferromagnetism with $T_{\rm C} \approx 85~\mathrm{K}$ originating from the localized $4f^7$ magnetic moments of Gd \cite{Ormaza2016}. The formation of multiple Weyl nodal-line-like band crossings in the occupied states has been reported by DFT calculations and ARPES \cite{Feng2019a}. It has been pointed out that the Weyl nodal line is protected by either spin-rotation U(1) symmetry or mirror symmetry $M_{\rm z}$ in the absence of SOC, whereas the inclusion of SOC and the magnetic configuration can lift these protections and open hybridization gaps \cite{Feng2019a}, leading to the finite Berry curvature and potentially strong anomalous Hall responses. Interestingly, although the temperature dependence of the electronic structure has not yet been systematically investigated, it has been proposed that spin splitting persists even above $T_{\rm C}$ \cite{Ormaza2016}, suggesting the realization of a spin-fluctuation-driven electronic state analogous to the Weyl state proposed for Eu-based compounds \cite{Ma2019,Milo2024}. Therefore, GdAg$_2$ provides an ideal platform for investigating how spin fluctuations influence SOC-induced band hybridization.

In this work, we systematically investigate the temperature evolution of the band structure of  GdAg$_2$/Ag(111) using ARPES, and demonstrate that localized spin moments generate spin splitting that persists to high temperatures, whereas SOC-induced hybridization emerges only in the presence of long-range ferromagnetic order. Furthermore, we discuss how the band-dependent quasiparticle scattering modifies the effective hybridization strength in conjunction with spin fluctuations, providing new insight into the evolution of the electronic structure near band crossings in systems with localized magnetic moments.

ARPES measurements were performed at Seoul National University using ScientaOmicron and SPECS hemispherical analyzers. Helium discharge lamps were used as the light source, providing predominantly linearly polarized photons with an energy of 21.2 eV selected by a diffraction grating. Polarization-dependent measurements with $s$- and $p$-polarized light were carried out using two experimental geometries. The overall energy resolution was set to 10 meV, and the sample temperature was varied between 10 and 250 K. Details of the sample preparation method and DFT calculations are provided in the Supplemental Information \cite{supple1}.

The surface alloy GdAg$_2$ is formed by depositing 1/3 monolayer of Gd onto an Ag(111) substrate \cite{Ormaza2016,Feng2019a,Correa2016,Fernandez2020}
. The topmost surface is characterized by a $(\sqrt{3} \times \sqrt{3})R30^\circ$ surface reconstruction [Fig. \ref{fig1}(b)], while lattice mismatch with the Ag substrate induces the periodic height variation, giving rise to a moiré pattern \cite{Correa2016}. Our LEED measurements also confirm the formation of a moiré pattern \cite{supple1}. 
Figure \ref{fig1}(c) shows the calculated band dispersion of monolayer GdAg$_2$ obtained from DFT without SOC. Spin-split electron and hole bands derived from hybridized Ag and Gd orbitals emerge due to the ferromagnetic order, and multiple band crossings are predicted near the Fermi level. When SOC is included, hybridization occurs between these bands, leading to the opening of SOC-induced gaps [Fig. \ref{fig1}(d)], as pointed out in previous studies \cite{Feng2019a}. Such a band structure is expected to produce a large anomalous Hall response if the transport property is measured.

\begin{figure}[t]
\begin{center}
\includegraphics[width=1\columnwidth]{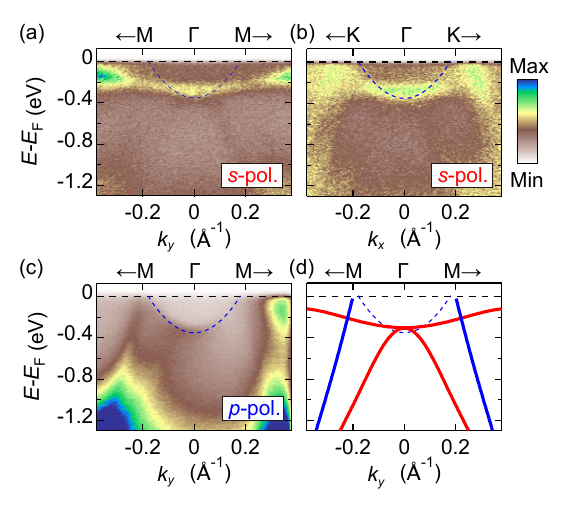}
\caption[]{(a) ARPES band maps measured along the $\bar{\rm M} - \Gamma - \bar{\rm M}$ direction with $s$-polarized light at $T=100$ K. The blue dotted curve denotes the edge of the projected bulk band continuum of Ag(111). (b) Same as (a), but measured along $\bar{\rm K} - \Gamma - \bar{\rm K}$. (c) Same as (a), but measured with $p$-polarized light. (d) Schematic illustration of the observed bands along $\bar{\rm M} - \Gamma - \bar{\rm M}$. Red and blue curves correspond to the bands that are expected to show majority spin polarization and minority spin polarization, respectively.}
\label{fig2}
\end{center}
\end{figure}

Figures  \ref{fig2}(a) and  \ref{fig2}(b) show ARPES band maps measured with $s$-polarized light at 100 K. Electron-like band dispersion with binding energies of approximately 0.3 eV is observed near the $\Gamma$ point. Comparison between the ${\rm M}-\Gamma-{\rm M}$ and ${\rm K}-\Gamma-{\rm K}$ directions reveals that the bands are nearly isotropic close to $\Gamma$, while strong anisotropy develops at larger momenta; in particular, pronounced photoemission intensity is observed at large momenta along the ${\rm K}-\Gamma-{\rm K}$ direction, indicating the contribution from different bands.

To investigate the band crossing, polarization-dependent measurements were carried out along the ${\rm M}-\Gamma-{\rm M}$ direction, where relatively simple dispersions are observed [Fig. \ref{fig2}(c)]. Remarkably, completely different intensity distributions are obtained for $s-$ and $p-$polarization. In particular, for $p-$polarization the intensity of the electron band vanishes, while clear intensity changes are observed at the edges of the L-projected bulk gap of the Ag substrate. Utilizing this polarization dependence, the dispersion of the hole band can be determined unambiguously. 
Figure \ref{fig2}(d) presents a schematic illustration of the experimentally observed band dispersions. By superimposing the bands measured with the two polarizations, we confirmed that the band dispersions are largely in agreement with the theoretical calculations. Importantly, a band crossing is confirmed around $k \sim \pm 0.2 {\rm \AA}^{-1}$, which is attributed to the nodal-line-derived feature highlighted in the right panel of Fig. \ref{fig1}(d). In contrast, band crossings predicted closer to the Fermi level by theory could not be resolved experimentally, possibly due to modifications of the band structure arising from enhanced hybridization with the Ag substrate or vertical displacements of the Gd atoms with respect to the topmost Ag plane, which may modify the band dispersions while keeping the band crossing discussed here \cite{Feng2019a}.

\begin{figure}[bth]
\begin{center}
\includegraphics[width=1\columnwidth]{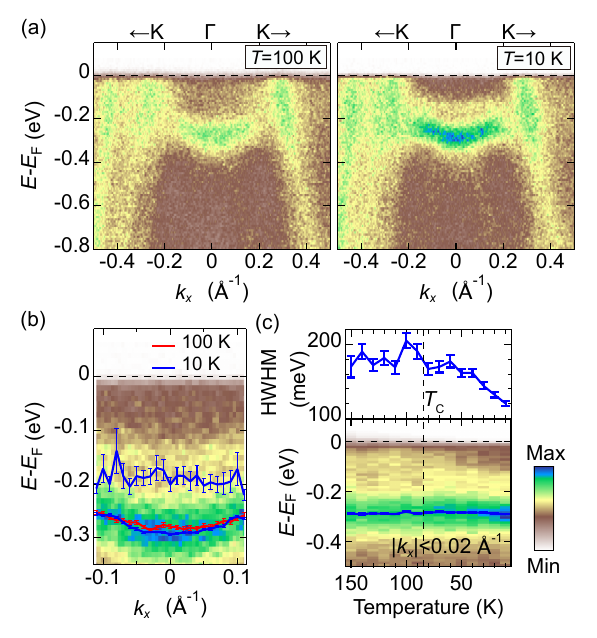}
\caption[]{(a) Temperature-dependent ARPES band maps taken at 100 K (left) and 10 K (right) along $\bar{\rm K} - \Gamma - \bar{\rm K}$ with $s$-polarized light. (b) Energy positions of the electron band and the subband-like feature obtained by two Lorentzian fits to the ARPES data taken at 100 K and 10 K, overlaid with the ARPES map taken at 10 K. The red and blue markers correspond to the results for 100 K and 10 K, respectively.  (c) Temperature-dependent energy positions (bottom) and spectral widths (top) at the $\Gamma$ point between 150 K and 10 K.}
\label{fig3}
\end{center}
\end{figure}

To examine the effect of the magnetic ordering, we proceeded to the temperature dependence of the band dispersion. Both electron- and hole-like features observed at high temperature remain visible at low temperature [Fig. \ref{fig3}(a)]. However, changes in the intensity distribution are observed near the Fermi level upon cooling.
A prominent main band with a binding energy of $\sim$0.3 eV around $\Gamma$ is observed regardless of the temperature. The energy positions of this band are nearly identical at 100 K and 10 K around $\Gamma$, as shown in Fig. \ref{fig3}(b). In contrast, while a broad photoemission intensity is observed near the Fermi level at 100 K, a weaker subband-like feature with a relatively flat dispersion becomes distinctly resolvable at low temperatures with a binding energy of $\sim$0.2 eV [Fig. \ref{fig3}(b)]. This subband is not predicted by DFT calculations and is therefore potentially associated with magnetic effects \cite{supple1}.
Further analysis of the detailed temperature-dependent ARPES results reveals that the main band responsible for the band crossing identified in Fig. \ref{fig2} exhibits temperature-dependent broadening at the $\Gamma$ point while maintaining a constant energy position, consistent with the spin-mixing mechanism [Fig. \ref{fig3}(c)]. Since the electron-like band is responsible for the formation of the band crossing illustrated in Fig. \ref{fig2}(d), the temperature independence of the energy position suggests the realization of spin-fluctuation-induced nodal-line-derived states in GdAg$_2$ above $T_{\rm C}$.
 
\begin{figure*}[t]
\begin{center}
\includegraphics[width=1\textwidth]{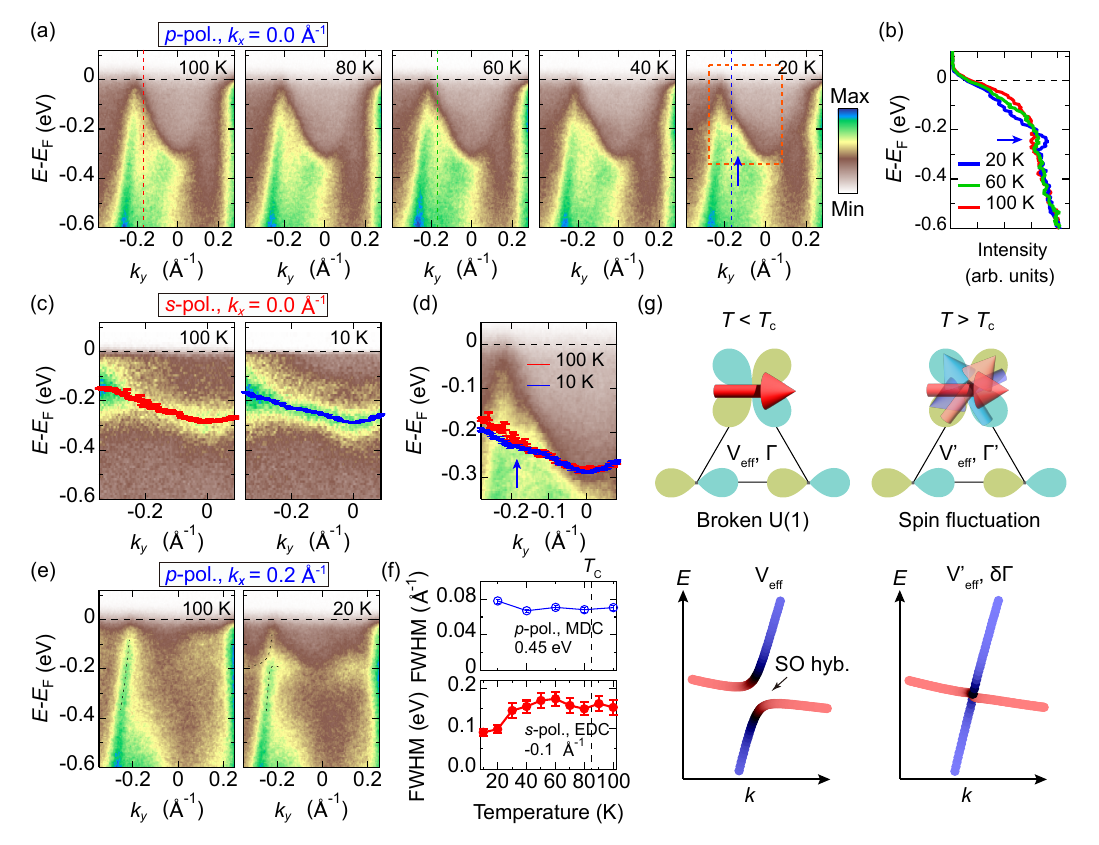}
\caption[]{(a) Temperature-dependent ARPES band maps measured with $p$-polarized light along the $\bar{\rm M} - \Gamma - \bar{\rm M}$ line at temperatures between 100 K and 20 K. (b) Energy distribution curves (EDCs) at the momentum indicated by the dashed lines in (a). The blue arrow highlights a peak that emerges at low temperature. (c) Magnified ARPES band maps measured with $s$-polarized light at 100 K and 10 K. Red and blue markers indicate the energy positions of the electron-like band at 100 K and 10 K, respectively.  (d) Energy positions of the electron-like band around the band crossing point at 100 K and 10 K, overlaid on the ARPES map measured with $p$-polarized light at 20 K [dotted red rectangle in (a)]. 
(e) ARPES band maps at $k_x = 0.2~\rm{\AA}^{-1}$ at 100 K and 20 K measured with $p$-polarized light. The black dotted lines are a guide to the eye. (f) Spectral widths extracted from momentum distribution curves (MDCs) for the hole-like bands and from EDCs for the electron-like band. 
(g) Schematic illustration of the spin-orbit-induced gap in GdAg$_2$/Ag(111), which is suppressed at higher temperatures due to spin decoherence and lifetime effects. The top and bottom panels represent real-space and momentum-space pictures, respectively. The color of the bands in the bottom panels indicates the right-eigenvector orbital character \cite{supple1}).}
\label{fig4}
\end{center}
\end{figure*}

Having confirmed the local moment magnetism manifested near the $\Gamma$ point, we next investigated the temperature dependence of the band structure in the vicinity of the band crossing to clarify the relationship between SOC-induced hybridization and magnetic fluctuations. The hole-like band measured with $p$-polarized light shows an approximately linear dispersion at 100 K, whereas a pronounced bending of the dispersion is observed as the temperature decreases [Fig. \ref{fig4}(a)]. This behavior is accompanied by a transfer of spectral weight from the vicinity of the Fermi level to higher binding energies [Fig. \ref{fig4}(b)]. Furthermore, the electron band observed with $s$-polarization also exhibits a small energy shift near the band crossing at low temperature [Fig. \ref{fig4}(c,d)]. These modifications of the spectral features indicate that band hybridization develops at low temperature, leading to changes in the dispersions near the crossing point and their orbital characters. 
The hybridization-induced intensity redistribution is more clearly observed in ARPES cuts taken away from the $\Gamma$ point, as illustrated in Fig. \ref{fig4}(e). 
Our observation evidences that the SOC-induced hybridization gap is formed below the ferromagnetic transition temperature, as predicted by DFT calculations. On the other hand, although the band splitting itself persists on the high-temperature side due to the local spin moments, the coherence of the SOC-induced hybridization is reduced as temperature increases.

The suppression of SOC-induced hybridization can be understood in terms of two cooperative effects. First, spin fluctuations reduce the coherence between spin components, effectively suppressing the SOC-induced hybridization matrix element. Second, band-dependent scattering introduces lifetime differences that reduce the hybridization gap through non-Hermitian effects. To account for both the reduced hybridization and the band-dependent linewidths within a unified framework, we introduce an effective two-orbital Hamiltonian, which can be written as \cite{Bentmann2012,Zhen2015,Eleuch2017,Kozii2024}
\begin{equation}
H_{\mathrm{eff}}(k)
=
\begin{pmatrix}
\varepsilon_1(k) & V_{\mathrm{eff}} \\
V_{\mathrm{eff}} & \varepsilon_2(k)
\end{pmatrix}
-
i
\begin{pmatrix}
\Gamma_1 & 0 \\
0 & \Gamma_2
\end{pmatrix},
\end{equation}
where $\varepsilon_{1,2}(k)$ denote the dispersions of the electron and hole bands, respectively, $V_{\mathrm{eff}}$ represents the SOC-induced interband hybridization, and $\Gamma_{1,2}$ correspond to the scattering rates of each band. The eigenenergies are then given by
\begin{equation}
E_{\pm}(k)
=
\frac{\varepsilon_1 + \varepsilon_2}{2}
-
i \bar{\Gamma}
\pm
\sqrt{
\left(
\frac{\varepsilon_1 - \varepsilon_2}{2}
-
i \delta\Gamma
\right)^2
+ V_{\mathrm{eff}}^2
}.
\end{equation}
, where $\bar{\Gamma}=(\Gamma_1 + \Gamma_2)/2$ and $\delta\Gamma=(\Gamma_1 - \Gamma_2)/2$ denote the average and difference of the scattering rates, respectively.
In particular, at the band crossing point $\varepsilon_1 = \varepsilon_2$, the splitting in the real part reads
\begin{equation}
E_{\mathrm{gap}}
=
2\,\mathrm{Re}
\sqrt{
V_{\mathrm{eff}}^2 - \delta\Gamma^2
}.
\end{equation}
Thus, while the SOC-induced hybridization $V_{\mathrm{eff}}$ sets the upper bound of the gap, the presence of a lifetime difference $\delta\Gamma$ between the two orbitals reduces the magnitude of the hybridization gap, and can eventually lead to a crossing-like spectrum despite finite hybridization when $\delta\Gamma$ exceeds $V_{\rm eff}$.

In GdAg$_2$, the SOC interaction, $V_{\mathrm{eff}} \sim \lambda \mathbf{L} \cdot \mathbf{S}$ leads to the well-defined hybridization at the lowest temperature. At higher temperatures, however, the direction of localized magnetization fluctuates in time, reducing the coherence of SOC-induced hybridization and resulting in a suppressed effective hybridization strength $V'_{\mathrm{eff}}$. Furthermore, the electron band exhibits a pronounced increase of $\sim 60~{\rm meV}$ in the full width at half maximum (FWHM) as the temperature increases from 20 to 60~K, indicating enhanced scattering and a shortened quasiparticle lifetime, whereas the hole band shows no significant change in linewidth [Fig. \ref{fig4}(f)]. Assuming a Lorentzian spectral function with FWHM $= 2\Gamma$, this corresponds to a change in the scattering rate of $\sim 30~\mathrm{meV}$ for the electron band, implying that the band-dependent scattering rate difference develops by $\sim 15~\mathrm{meV}$. These results indicate that non-Hermitian effects arising from band-dependent damping play a key role in the temperature evolution of SOC-induced hybridization.
Consequently, GdAg$_2$ can be regarded as a unique platform in which the SOC hybridization is controlled by the combined effects of reduced effective SOC coupling due to spin fluctuations and band-dependent scattering rates as illustrated in Fig. \ref{fig4}(g).

Our results suggest that the SOC-induced band hybridization in GdAg$_2$ is not a fixed property, but can be efficiently tuned by magnetic fluctuations. The strong sensitivity of SOC hybridization to spin fluctuations provides a route to control the Berry curvature distribution and related transport properties, such as the anomalous Hall effect, by tuning the magnetic state via temperature or magnetic field. In contrast to previously demonstrated SOC hybridization effects governed by the spin orientation \cite{Cheng2023,Lou2024}, the use of spin fluctuations allows easy access to continuous modulation of the hybridization strength. 
Moreover, the spectral evolution at the lowest temperatures suggests that magnetic materials provide a natural platform for exploring non-Hermitian band physics. The coexistence of SOC-induced hybridization and band-dependent damping allows the system to approach a regime where the hybridization strength $V_{\mathrm{eff}}$ and the lifetime difference $\delta\Gamma$ become comparable, a necessary condition for investigating non-Hermitian band crossings \cite{Zhen2015,Eleuch2017,Kozii2024}. Since the band-dependent damping is widely observed in magnetic materials \cite{Jo2021,Hahn2021}, this scheme provides a promising route for experimentally probing non-Hermitian band structures in correlated electron systems \cite{Zhen2015,Eleuch2017,Bergholtz2021,Cayao2023,Kozii2024,Yoshida2018,Kawabata2019,Bergholtz2019,Nagai2020,CongLi2025}.

In conclusion, by analyzing the temperature and polarization dependence of the electronic band structure of the surface ferromagnet GdAg$_2$/Ag(111), we demonstrate that strong localized Gd moments induce band splitting even above $T_{\rm C}$, stabilizing nodal-line-like band crossings via spin fluctuations. Meanwhile, a spectral weight redistribution is observed near the band crossing, indicating that SOC-induced band hybridization is stabilized only in the presence of long-range magnetic order. 
This temperature dependence can be understood by incorporating spin decoherence, which renormalizes the effective SOC hybridization, and band-dependent lifetime effects, which introduce non-Hermitian contributions. Our findings therefore highlight magnetic ordering and its associated spin fluctuations as a key control parameter for engineering SOC-driven band hybridization, topological electronic responses, and non-Hermitian band crossings.

\section{Acknowledgements}
We thank Sangjae Lee and Soonsang Huh for helpful discussions.
This work was supported by the Institute for Basic Science in the Republic of Korea (Grant Nos. IBS-R009-G2) and by the National Research
Foundation of Korea (NRF) (Grant Nos. RS-2023-00258359 and NRF-
2022R1A3B1077234). Part of this work was also supported by the Institute of Applied Physics, Seoul National University.
R.N. acknowledges the support from the Ministry of Education (MOE) in Taiwan through the Yushan Fellow Program under grant No. 114V1072-1 and the Higher Education Sprout Project under Grants No. 114L900802 and from the National Science and Technology Council (NSTC) of Taiwan under Grants No. NSTC 114-2628-M-002-011.





\begin{thebibliography}{60}%
\makeatletter
\providecommand \@ifxundefined [1]{%
 \@ifx{#1\undefined}
}%
\providecommand \@ifnum [1]{%
 \ifnum #1\expandafter \@firstoftwo
 \else \expandafter \@secondoftwo
 \fi
}%
\providecommand \@ifx [1]{%
 \ifx #1\expandafter \@firstoftwo
 \else \expandafter \@secondoftwo
 \fi
}%
\providecommand \natexlab [1]{#1}%
\providecommand \enquote  [1]{``#1''}%
\providecommand \bibnamefont  [1]{#1}%
\providecommand \bibfnamefont [1]{#1}%
\providecommand \citenamefont [1]{#1}%
\providecommand \href@noop [0]{\@secondoftwo}%
\providecommand \href [0]{\begingroup \@sanitize@url \@href}%
\providecommand \@href[1]{\@@startlink{#1}\@@href}%
\providecommand \@@href[1]{\endgroup#1\@@endlink}%
\providecommand \@sanitize@url [0]{\catcode `\\12\catcode `\$12\catcode `\&12\catcode `\#12\catcode `\^12\catcode `\_12\catcode `\%12\relax}%
\providecommand \@@startlink[1]{}%
\providecommand \@@endlink[0]{}%
\providecommand \url  [0]{\begingroup\@sanitize@url \@url }%
\providecommand \@url [1]{\endgroup\@href {#1}{\urlprefix }}%
\providecommand \urlprefix  [0]{URL }%
\providecommand \Eprint [0]{\href }%
\providecommand \doibase [0]{https://doi.org/}%
\providecommand \selectlanguage [0]{\@gobble}%
\providecommand \bibinfo  [0]{\@secondoftwo}%
\providecommand \bibfield  [0]{\@secondoftwo}%
\providecommand \translation [1]{[#1]}%
\providecommand \BibitemOpen [0]{}%
\providecommand \bibitemStop [0]{}%
\providecommand \bibitemNoStop [0]{.\EOS\space}%
\providecommand \EOS [0]{\spacefactor3000\relax}%
\providecommand \BibitemShut  [1]{\csname bibitem#1\endcsname}%
\let\auto@bib@innerbib\@empty
\bibitem [{\citenamefont {Tokura}\ \emph {et~al.}(2019)\citenamefont {Tokura}, \citenamefont {Yasuda},\ and\ \citenamefont {Tsukazaki}}]{Tokura2019}%
  \BibitemOpen
  \bibfield  {author} {\bibinfo {author} {\bibfnamefont {Y.}~\bibnamefont {Tokura}}, \bibinfo {author} {\bibfnamefont {K.}~\bibnamefont {Yasuda}},\ and\ \bibinfo {author} {\bibfnamefont {A.}~\bibnamefont {Tsukazaki}},\ }\href {https://doi.org/10.1038/s42254-018-0011-5} {\bibfield  {journal} {\bibinfo  {journal} {Nature Reviews Physics}\ }\textbf {\bibinfo {volume} {1}},\ \bibinfo {pages} {126} (\bibinfo {year} {2019})}\BibitemShut {NoStop}%
\bibitem [{\citenamefont {Chang}\ \emph {et~al.}(2023)\citenamefont {Chang}, \citenamefont {Liu},\ and\ \citenamefont {MacDonald}}]{Chang2023}%
  \BibitemOpen
  \bibfield  {author} {\bibinfo {author} {\bibfnamefont {C.-Z.}\ \bibnamefont {Chang}}, \bibinfo {author} {\bibfnamefont {C.-X.}\ \bibnamefont {Liu}},\ and\ \bibinfo {author} {\bibfnamefont {A.~H.}\ \bibnamefont {MacDonald}},\ }\href {https://doi.org/10.1103/RevModPhys.95.011002} {\bibfield  {journal} {\bibinfo  {journal} {Reviews of Modern Physics}\ }\textbf {\bibinfo {volume} {95}},\ \bibinfo {pages} {011002} (\bibinfo {year} {2023})}\BibitemShut {NoStop}%
\bibitem [{\citenamefont {Chang}\ \emph {et~al.}(2013)\citenamefont {Chang}, \citenamefont {Zhang}, \citenamefont {Feng}, \citenamefont {Shen},\ and\ \citenamefont {Zhang}}]{Chang2013a}%
  \BibitemOpen
  \bibfield  {author} {\bibinfo {author} {\bibfnamefont {C.}~\bibnamefont {Chang}}, \bibinfo {author} {\bibfnamefont {J.}~\bibnamefont {Zhang}}, \bibinfo {author} {\bibfnamefont {X.}~\bibnamefont {Feng}}, \bibinfo {author} {\bibfnamefont {J.}~\bibnamefont {Shen}},\ and\ \bibinfo {author} {\bibfnamefont {Z.}~\bibnamefont {Zhang}},\ }\href {http://www.sciencemag.org/content/340/6129/167.short} {\bibfield  {journal} {\bibinfo  {journal} {Science}\ }\textbf {\bibinfo {volume} {15}},\ \bibinfo {pages} {167} (\bibinfo {year} {2013})}\BibitemShut {NoStop}%
\bibitem [{\citenamefont {Checkelsky}\ \emph {et~al.}(2014)\citenamefont {Checkelsky}, \citenamefont {Yoshimi}, \citenamefont {Tsukazaki}, \citenamefont {Takahashi}, \citenamefont {Kozuka}, \citenamefont {Falson}, \citenamefont {Kawasaki},\ and\ \citenamefont {Tokura}}]{Checkelsky2014}%
  \BibitemOpen
  \bibfield  {author} {\bibinfo {author} {\bibfnamefont {J.~G.}\ \bibnamefont {Checkelsky}}, \bibinfo {author} {\bibfnamefont {R.}~\bibnamefont {Yoshimi}}, \bibinfo {author} {\bibfnamefont {A.}~\bibnamefont {Tsukazaki}}, \bibinfo {author} {\bibfnamefont {K.~S.}\ \bibnamefont {Takahashi}}, \bibinfo {author} {\bibfnamefont {Y.}~\bibnamefont {Kozuka}}, \bibinfo {author} {\bibfnamefont {J.}~\bibnamefont {Falson}}, \bibinfo {author} {\bibfnamefont {M.}~\bibnamefont {Kawasaki}},\ and\ \bibinfo {author} {\bibfnamefont {Y.}~\bibnamefont {Tokura}},\ }\href {https://doi.org/10.1038/nphys3053} {\bibfield  {journal} {\bibinfo  {journal} {Nature Physics}\ }\textbf {\bibinfo {volume} {10}},\ \bibinfo {pages} {731} (\bibinfo {year} {2014})}\BibitemShut {NoStop}%
\bibitem [{\citenamefont {Otrokov}\ \emph {et~al.}(2019)\citenamefont {Otrokov}, \citenamefont {Klimovskikh}, \citenamefont {Bentmann}, \citenamefont {Estyunin}, \citenamefont {Zeugner}, \citenamefont {Aliev}, \citenamefont {Gaß}, \citenamefont {Wolter}, \citenamefont {Koroleva}, \citenamefont {Shikin}, \citenamefont {Blanco-Rey}, \citenamefont {Hoffmann}, \citenamefont {Rusinov}, \citenamefont {Vyazovskaya}, \citenamefont {Eremeev}, \citenamefont {Koroteev}, \citenamefont {Kuznetsov}, \citenamefont {Freyse}, \citenamefont {Sánchez-Barriga}, \citenamefont {Amiraslanov}, \citenamefont {Babanly}, \citenamefont {Mamedov}, \citenamefont {Abdullayev}, \citenamefont {Zverev}, \citenamefont {Alfonsov}, \citenamefont {Kataev}, \citenamefont {Büchner}, \citenamefont {Schwier}, \citenamefont {Kumar}, \citenamefont {Kimura}, \citenamefont {Petaccia}, \citenamefont {Santo}, \citenamefont {Vidal}, \citenamefont {Schatz}, \citenamefont {Kißner}, \citenamefont {Ünzelmann}, \citenamefont {Min}, \citenamefont {Moser},
  \citenamefont {Peixoto}, \citenamefont {Reinert}, \citenamefont {Ernst}, \citenamefont {Echenique}, \citenamefont {Isaeva},\ and\ \citenamefont {Chulkov}}]{Otrokov2019}%
  \BibitemOpen
  \bibfield  {author} {\bibinfo {author} {\bibfnamefont {M.~M.}\ \bibnamefont {Otrokov}}, \bibinfo {author} {\bibfnamefont {I.~I.}\ \bibnamefont {Klimovskikh}}, \bibinfo {author} {\bibfnamefont {H.}~\bibnamefont {Bentmann}}, \bibinfo {author} {\bibfnamefont {D.}~\bibnamefont {Estyunin}}, \bibinfo {author} {\bibfnamefont {A.}~\bibnamefont {Zeugner}}, \bibinfo {author} {\bibfnamefont {Z.~S.}\ \bibnamefont {Aliev}}, \bibinfo {author} {\bibfnamefont {S.}~\bibnamefont {Gaß}}, \bibinfo {author} {\bibfnamefont {A.~U.~B.}\ \bibnamefont {Wolter}}, \bibinfo {author} {\bibfnamefont {A.~V.}\ \bibnamefont {Koroleva}}, \bibinfo {author} {\bibfnamefont {A.~M.}\ \bibnamefont {Shikin}}, \bibinfo {author} {\bibfnamefont {M.}~\bibnamefont {Blanco-Rey}}, \bibinfo {author} {\bibfnamefont {M.}~\bibnamefont {Hoffmann}}, \bibinfo {author} {\bibfnamefont {I.~P.}\ \bibnamefont {Rusinov}}, \bibinfo {author} {\bibfnamefont {A.~Y.}\ \bibnamefont {Vyazovskaya}}, \bibinfo {author} {\bibfnamefont {S.~V.}\ \bibnamefont {Eremeev}}, \bibinfo
  {author} {\bibfnamefont {Y.~M.}\ \bibnamefont {Koroteev}}, \bibinfo {author} {\bibfnamefont {V.~M.}\ \bibnamefont {Kuznetsov}}, \bibinfo {author} {\bibfnamefont {F.}~\bibnamefont {Freyse}}, \bibinfo {author} {\bibfnamefont {J.}~\bibnamefont {Sánchez-Barriga}}, \bibinfo {author} {\bibfnamefont {I.~R.}\ \bibnamefont {Amiraslanov}}, \bibinfo {author} {\bibfnamefont {M.~B.}\ \bibnamefont {Babanly}}, \bibinfo {author} {\bibfnamefont {N.~T.}\ \bibnamefont {Mamedov}}, \bibinfo {author} {\bibfnamefont {N.~A.}\ \bibnamefont {Abdullayev}}, \bibinfo {author} {\bibfnamefont {V.~N.}\ \bibnamefont {Zverev}}, \bibinfo {author} {\bibfnamefont {A.}~\bibnamefont {Alfonsov}}, \bibinfo {author} {\bibfnamefont {V.}~\bibnamefont {Kataev}}, \bibinfo {author} {\bibfnamefont {B.}~\bibnamefont {Büchner}}, \bibinfo {author} {\bibfnamefont {E.~F.}\ \bibnamefont {Schwier}}, \bibinfo {author} {\bibfnamefont {S.}~\bibnamefont {Kumar}}, \bibinfo {author} {\bibfnamefont {A.}~\bibnamefont {Kimura}}, \bibinfo {author} {\bibfnamefont
  {L.}~\bibnamefont {Petaccia}}, \bibinfo {author} {\bibfnamefont {G.~D.}\ \bibnamefont {Santo}}, \bibinfo {author} {\bibfnamefont {R.~C.}\ \bibnamefont {Vidal}}, \bibinfo {author} {\bibfnamefont {S.}~\bibnamefont {Schatz}}, \bibinfo {author} {\bibfnamefont {K.}~\bibnamefont {Kißner}}, \bibinfo {author} {\bibfnamefont {M.}~\bibnamefont {Ünzelmann}}, \bibinfo {author} {\bibfnamefont {C.~H.}\ \bibnamefont {Min}}, \bibinfo {author} {\bibfnamefont {S.}~\bibnamefont {Moser}}, \bibinfo {author} {\bibfnamefont {T.~R.~F.}\ \bibnamefont {Peixoto}}, \bibinfo {author} {\bibfnamefont {F.}~\bibnamefont {Reinert}}, \bibinfo {author} {\bibfnamefont {A.}~\bibnamefont {Ernst}}, \bibinfo {author} {\bibfnamefont {P.~M.}\ \bibnamefont {Echenique}}, \bibinfo {author} {\bibfnamefont {A.}~\bibnamefont {Isaeva}},\ and\ \bibinfo {author} {\bibfnamefont {E.~V.}\ \bibnamefont {Chulkov}},\ }\href {https://doi.org/10.1038/s41586-019-1840-9} {\bibfield  {journal} {\bibinfo  {journal} {Nature}\ }\textbf {\bibinfo {volume} {576}},\
  \bibinfo {pages} {416} (\bibinfo {year} {2019})}\BibitemShut {NoStop}%
\bibitem [{\citenamefont {Gong}\ \emph {et~al.}(2019)\citenamefont {Gong}, \citenamefont {Guo}, \citenamefont {Li}, \citenamefont {Zhu}, \citenamefont {Liao}, \citenamefont {Liu}, \citenamefont {Zhang}, \citenamefont {Gu}, \citenamefont {Tang}, \citenamefont {Feng}, \citenamefont {Zhang}, \citenamefont {Li}, \citenamefont {Song}, \citenamefont {Wang}, \citenamefont {Yu}, \citenamefont {Chen}, \citenamefont {Wang}, \citenamefont {Yao}, \citenamefont {Duan}, \citenamefont {Xu}, \citenamefont {Zhang}, \citenamefont {Ma}, \citenamefont {Xue},\ and\ \citenamefont {He}}]{Gong2018a}%
  \BibitemOpen
  \bibfield  {author} {\bibinfo {author} {\bibfnamefont {Y.}~\bibnamefont {Gong}}, \bibinfo {author} {\bibfnamefont {J.}~\bibnamefont {Guo}}, \bibinfo {author} {\bibfnamefont {J.}~\bibnamefont {Li}}, \bibinfo {author} {\bibfnamefont {K.}~\bibnamefont {Zhu}}, \bibinfo {author} {\bibfnamefont {M.}~\bibnamefont {Liao}}, \bibinfo {author} {\bibfnamefont {X.}~\bibnamefont {Liu}}, \bibinfo {author} {\bibfnamefont {Q.}~\bibnamefont {Zhang}}, \bibinfo {author} {\bibfnamefont {L.}~\bibnamefont {Gu}}, \bibinfo {author} {\bibfnamefont {L.}~\bibnamefont {Tang}}, \bibinfo {author} {\bibfnamefont {X.}~\bibnamefont {Feng}}, \bibinfo {author} {\bibfnamefont {D.}~\bibnamefont {Zhang}}, \bibinfo {author} {\bibfnamefont {W.}~\bibnamefont {Li}}, \bibinfo {author} {\bibfnamefont {C.}~\bibnamefont {Song}}, \bibinfo {author} {\bibfnamefont {L.}~\bibnamefont {Wang}}, \bibinfo {author} {\bibfnamefont {P.}~\bibnamefont {Yu}}, \bibinfo {author} {\bibfnamefont {X.}~\bibnamefont {Chen}}, \bibinfo {author} {\bibfnamefont {Y.}~\bibnamefont
  {Wang}}, \bibinfo {author} {\bibfnamefont {H.}~\bibnamefont {Yao}}, \bibinfo {author} {\bibfnamefont {W.}~\bibnamefont {Duan}}, \bibinfo {author} {\bibfnamefont {Y.}~\bibnamefont {Xu}}, \bibinfo {author} {\bibfnamefont {S.-C.}\ \bibnamefont {Zhang}}, \bibinfo {author} {\bibfnamefont {X.}~\bibnamefont {Ma}}, \bibinfo {author} {\bibfnamefont {Q.-K.}\ \bibnamefont {Xue}},\ and\ \bibinfo {author} {\bibfnamefont {K.}~\bibnamefont {He}},\ }\href {https://doi.org/10.1088/0256-307X/36/7/076801} {\bibfield  {journal} {\bibinfo  {journal} {Chinese Physics Letters}\ }\textbf {\bibinfo {volume} {36}},\ \bibinfo {pages} {076801} (\bibinfo {year} {2019})}\BibitemShut {NoStop}%
\bibitem [{\citenamefont {Deng}\ \emph {et~al.}(2020)\citenamefont {Deng}, \citenamefont {Yu}, \citenamefont {Shi}, \citenamefont {Guo}, \citenamefont {Xu}, \citenamefont {Wang}, \citenamefont {Chen},\ and\ \citenamefont {Zhang}}]{Deng2020}%
  \BibitemOpen
  \bibfield  {author} {\bibinfo {author} {\bibfnamefont {Y.}~\bibnamefont {Deng}}, \bibinfo {author} {\bibfnamefont {Y.}~\bibnamefont {Yu}}, \bibinfo {author} {\bibfnamefont {M.~Z.}\ \bibnamefont {Shi}}, \bibinfo {author} {\bibfnamefont {Z.}~\bibnamefont {Guo}}, \bibinfo {author} {\bibfnamefont {Z.}~\bibnamefont {Xu}}, \bibinfo {author} {\bibfnamefont {J.}~\bibnamefont {Wang}}, \bibinfo {author} {\bibfnamefont {X.~H.}\ \bibnamefont {Chen}},\ and\ \bibinfo {author} {\bibfnamefont {Y.}~\bibnamefont {Zhang}},\ }\href {https://doi.org/10.1126/science.aax8156} {\bibfield  {journal} {\bibinfo  {journal} {Science}\ }\textbf {\bibinfo {volume} {367}},\ \bibinfo {pages} {895} (\bibinfo {year} {2020})}\BibitemShut {NoStop}%
\bibitem [{\citenamefont {Kuroda}\ \emph {et~al.}(2017)\citenamefont {Kuroda}, \citenamefont {Tomita}, \citenamefont {Suzuki}, \citenamefont {Bareille}, \citenamefont {Nugroho}, \citenamefont {Goswami}, \citenamefont {Ochi}, \citenamefont {Ikhlas}, \citenamefont {Nakayama}, \citenamefont {Akebi}, \citenamefont {Noguchi}, \citenamefont {Ishii}, \citenamefont {Inami}, \citenamefont {Ono}, \citenamefont {Kumigashira}, \citenamefont {Varykhalov}, \citenamefont {Muro}, \citenamefont {Koretsune}, \citenamefont {Arita}, \citenamefont {Shin}, \citenamefont {Kondo},\ and\ \citenamefont {Nakatsuji}}]{Kuroda2017b}%
  \BibitemOpen
  \bibfield  {author} {\bibinfo {author} {\bibfnamefont {K.}~\bibnamefont {Kuroda}}, \bibinfo {author} {\bibfnamefont {T.}~\bibnamefont {Tomita}}, \bibinfo {author} {\bibfnamefont {M.-T.}\ \bibnamefont {Suzuki}}, \bibinfo {author} {\bibfnamefont {C.}~\bibnamefont {Bareille}}, \bibinfo {author} {\bibfnamefont {A.~A.}\ \bibnamefont {Nugroho}}, \bibinfo {author} {\bibfnamefont {P.}~\bibnamefont {Goswami}}, \bibinfo {author} {\bibfnamefont {M.}~\bibnamefont {Ochi}}, \bibinfo {author} {\bibfnamefont {M.}~\bibnamefont {Ikhlas}}, \bibinfo {author} {\bibfnamefont {M.}~\bibnamefont {Nakayama}}, \bibinfo {author} {\bibfnamefont {S.}~\bibnamefont {Akebi}}, \bibinfo {author} {\bibfnamefont {R.}~\bibnamefont {Noguchi}}, \bibinfo {author} {\bibfnamefont {R.}~\bibnamefont {Ishii}}, \bibinfo {author} {\bibfnamefont {N.}~\bibnamefont {Inami}}, \bibinfo {author} {\bibfnamefont {K.}~\bibnamefont {Ono}}, \bibinfo {author} {\bibfnamefont {H.}~\bibnamefont {Kumigashira}}, \bibinfo {author} {\bibfnamefont {A.}~\bibnamefont
  {Varykhalov}}, \bibinfo {author} {\bibfnamefont {T.}~\bibnamefont {Muro}}, \bibinfo {author} {\bibfnamefont {T.}~\bibnamefont {Koretsune}}, \bibinfo {author} {\bibfnamefont {R.}~\bibnamefont {Arita}}, \bibinfo {author} {\bibfnamefont {S.}~\bibnamefont {Shin}}, \bibinfo {author} {\bibfnamefont {T.}~\bibnamefont {Kondo}},\ and\ \bibinfo {author} {\bibfnamefont {S.}~\bibnamefont {Nakatsuji}},\ }\href {https://doi.org/10.1038/nmat4987} {\bibfield  {journal} {\bibinfo  {journal} {Nature Materials}\ }\textbf {\bibinfo {volume} {16}},\ \bibinfo {pages} {1090} (\bibinfo {year} {2017})}\BibitemShut {NoStop}%
\bibitem [{\citenamefont {Liu}\ \emph {et~al.}(2019)\citenamefont {Liu}, \citenamefont {Liang}, \citenamefont {Liu}, \citenamefont {Xu}, \citenamefont {Li}, \citenamefont {Chen}, \citenamefont {Pei}, \citenamefont {Shi}, \citenamefont {Mo}, \citenamefont {Dudin}, \citenamefont {Kim}, \citenamefont {Cacho}, \citenamefont {Li}, \citenamefont {Sun}, \citenamefont {Yang}, \citenamefont {Liu}, \citenamefont {Parkin}, \citenamefont {Felser},\ and\ \citenamefont {Chen}}]{Liu2019d}%
  \BibitemOpen
  \bibfield  {author} {\bibinfo {author} {\bibfnamefont {D.~F.}\ \bibnamefont {Liu}}, \bibinfo {author} {\bibfnamefont {A.~J.}\ \bibnamefont {Liang}}, \bibinfo {author} {\bibfnamefont {E.~K.}\ \bibnamefont {Liu}}, \bibinfo {author} {\bibfnamefont {Q.~N.}\ \bibnamefont {Xu}}, \bibinfo {author} {\bibfnamefont {Y.~W.}\ \bibnamefont {Li}}, \bibinfo {author} {\bibfnamefont {C.}~\bibnamefont {Chen}}, \bibinfo {author} {\bibfnamefont {D.}~\bibnamefont {Pei}}, \bibinfo {author} {\bibfnamefont {W.~J.}\ \bibnamefont {Shi}}, \bibinfo {author} {\bibfnamefont {S.~K.}\ \bibnamefont {Mo}}, \bibinfo {author} {\bibfnamefont {P.}~\bibnamefont {Dudin}}, \bibinfo {author} {\bibfnamefont {T.}~\bibnamefont {Kim}}, \bibinfo {author} {\bibfnamefont {C.}~\bibnamefont {Cacho}}, \bibinfo {author} {\bibfnamefont {G.}~\bibnamefont {Li}}, \bibinfo {author} {\bibfnamefont {Y.}~\bibnamefont {Sun}}, \bibinfo {author} {\bibfnamefont {L.~X.}\ \bibnamefont {Yang}}, \bibinfo {author} {\bibfnamefont {Z.~K.}\ \bibnamefont {Liu}}, \bibinfo {author}
  {\bibfnamefont {S.~S.~P.}\ \bibnamefont {Parkin}}, \bibinfo {author} {\bibfnamefont {C.}~\bibnamefont {Felser}},\ and\ \bibinfo {author} {\bibfnamefont {Y.~L.}\ \bibnamefont {Chen}},\ }\href {https://doi.org/10.1126/science.aav2873} {\bibfield  {journal} {\bibinfo  {journal} {Science}\ }\textbf {\bibinfo {volume} {365}},\ \bibinfo {pages} {1282} (\bibinfo {year} {2019})}\BibitemShut {NoStop}%
\bibitem [{\citenamefont {Morali}\ \emph {et~al.}(2019)\citenamefont {Morali}, \citenamefont {Batabyal}, \citenamefont {Nag}, \citenamefont {Liu}, \citenamefont {Xu}, \citenamefont {Sun}, \citenamefont {Yan}, \citenamefont {Felser}, \citenamefont {Avraham},\ and\ \citenamefont {Beidenkopf}}]{Morali2019}%
  \BibitemOpen
  \bibfield  {author} {\bibinfo {author} {\bibfnamefont {N.}~\bibnamefont {Morali}}, \bibinfo {author} {\bibfnamefont {R.}~\bibnamefont {Batabyal}}, \bibinfo {author} {\bibfnamefont {P.~K.}\ \bibnamefont {Nag}}, \bibinfo {author} {\bibfnamefont {E.}~\bibnamefont {Liu}}, \bibinfo {author} {\bibfnamefont {Q.}~\bibnamefont {Xu}}, \bibinfo {author} {\bibfnamefont {Y.}~\bibnamefont {Sun}}, \bibinfo {author} {\bibfnamefont {B.}~\bibnamefont {Yan}}, \bibinfo {author} {\bibfnamefont {C.}~\bibnamefont {Felser}}, \bibinfo {author} {\bibfnamefont {N.}~\bibnamefont {Avraham}},\ and\ \bibinfo {author} {\bibfnamefont {H.}~\bibnamefont {Beidenkopf}},\ }\href {https://doi.org/10.1126/science.aav2334} {\bibfield  {journal} {\bibinfo  {journal} {Science}\ }\textbf {\bibinfo {volume} {365}},\ \bibinfo {pages} {1286} (\bibinfo {year} {2019})}\BibitemShut {NoStop}%
\bibitem [{\citenamefont {Belopolski}\ \emph {et~al.}(2019)\citenamefont {Belopolski}, \citenamefont {Manna}, \citenamefont {Sanchez}, \citenamefont {Chang}, \citenamefont {Ernst}, \citenamefont {Yin}, \citenamefont {Zhang}, \citenamefont {Cochran}, \citenamefont {Shumiya}, \citenamefont {Zheng}, \citenamefont {Singh}, \citenamefont {Bian}, \citenamefont {Multer}, \citenamefont {Litskevich}, \citenamefont {Zhou}, \citenamefont {Huang}, \citenamefont {Wang}, \citenamefont {Chang}, \citenamefont {Xu}, \citenamefont {Bansil}, \citenamefont {Felser}, \citenamefont {Lin},\ and\ \citenamefont {Hasan}}]{Belopolski2019}%
  \BibitemOpen
  \bibfield  {author} {\bibinfo {author} {\bibfnamefont {I.}~\bibnamefont {Belopolski}}, \bibinfo {author} {\bibfnamefont {K.}~\bibnamefont {Manna}}, \bibinfo {author} {\bibfnamefont {D.~S.}\ \bibnamefont {Sanchez}}, \bibinfo {author} {\bibfnamefont {G.}~\bibnamefont {Chang}}, \bibinfo {author} {\bibfnamefont {B.}~\bibnamefont {Ernst}}, \bibinfo {author} {\bibfnamefont {J.}~\bibnamefont {Yin}}, \bibinfo {author} {\bibfnamefont {S.~S.}\ \bibnamefont {Zhang}}, \bibinfo {author} {\bibfnamefont {T.}~\bibnamefont {Cochran}}, \bibinfo {author} {\bibfnamefont {N.}~\bibnamefont {Shumiya}}, \bibinfo {author} {\bibfnamefont {H.}~\bibnamefont {Zheng}}, \bibinfo {author} {\bibfnamefont {B.}~\bibnamefont {Singh}}, \bibinfo {author} {\bibfnamefont {G.}~\bibnamefont {Bian}}, \bibinfo {author} {\bibfnamefont {D.}~\bibnamefont {Multer}}, \bibinfo {author} {\bibfnamefont {M.}~\bibnamefont {Litskevich}}, \bibinfo {author} {\bibfnamefont {X.}~\bibnamefont {Zhou}}, \bibinfo {author} {\bibfnamefont {S.-M.}\ \bibnamefont {Huang}},
  \bibinfo {author} {\bibfnamefont {B.}~\bibnamefont {Wang}}, \bibinfo {author} {\bibfnamefont {T.-R.}\ \bibnamefont {Chang}}, \bibinfo {author} {\bibfnamefont {S.-Y.}\ \bibnamefont {Xu}}, \bibinfo {author} {\bibfnamefont {A.}~\bibnamefont {Bansil}}, \bibinfo {author} {\bibfnamefont {C.}~\bibnamefont {Felser}}, \bibinfo {author} {\bibfnamefont {H.}~\bibnamefont {Lin}},\ and\ \bibinfo {author} {\bibfnamefont {M.~Z.}\ \bibnamefont {Hasan}},\ }\href {https://doi.org/10.1126/science.aav2327} {\bibfield  {journal} {\bibinfo  {journal} {Science}\ }\textbf {\bibinfo {volume} {365}},\ \bibinfo {pages} {1278} (\bibinfo {year} {2019})}\BibitemShut {NoStop}%
\bibitem [{\citenamefont {Xu}\ \emph {et~al.}(2018)\citenamefont {Xu}, \citenamefont {Liu}, \citenamefont {Shi}, \citenamefont {Muechler}, \citenamefont {Gayles}, \citenamefont {Felser},\ and\ \citenamefont {Sun}}]{Xu2018aa}%
  \BibitemOpen
  \bibfield  {author} {\bibinfo {author} {\bibfnamefont {Q.}~\bibnamefont {Xu}}, \bibinfo {author} {\bibfnamefont {E.}~\bibnamefont {Liu}}, \bibinfo {author} {\bibfnamefont {W.}~\bibnamefont {Shi}}, \bibinfo {author} {\bibfnamefont {L.}~\bibnamefont {Muechler}}, \bibinfo {author} {\bibfnamefont {J.}~\bibnamefont {Gayles}}, \bibinfo {author} {\bibfnamefont {C.}~\bibnamefont {Felser}},\ and\ \bibinfo {author} {\bibfnamefont {Y.}~\bibnamefont {Sun}},\ }\href {https://doi.org/10.1103/PhysRevB.97.235416} {\bibfield  {journal} {\bibinfo  {journal} {Physical Review B}\ }\textbf {\bibinfo {volume} {97}},\ \bibinfo {pages} {235416} (\bibinfo {year} {2018})}\BibitemShut {NoStop}%
\bibitem [{\citenamefont {Liu}\ \emph {et~al.}(2018)\citenamefont {Liu}, \citenamefont {Sun}, \citenamefont {Kumar}, \citenamefont {Muechler}, \citenamefont {Sun}, \citenamefont {Jiao}, \citenamefont {Yang}, \citenamefont {Liu}, \citenamefont {Liang}, \citenamefont {Xu}, \citenamefont {Kroder}, \citenamefont {Süß}, \citenamefont {Borrmann}, \citenamefont {Shekhar}, \citenamefont {Wang}, \citenamefont {Xi}, \citenamefont {Wang}, \citenamefont {Schnelle}, \citenamefont {Wirth}, \citenamefont {Chen}, \citenamefont {Goennenwein},\ and\ \citenamefont {Felser}}]{Liu2018aa}%
  \BibitemOpen
  \bibfield  {author} {\bibinfo {author} {\bibfnamefont {E.}~\bibnamefont {Liu}}, \bibinfo {author} {\bibfnamefont {Y.}~\bibnamefont {Sun}}, \bibinfo {author} {\bibfnamefont {N.}~\bibnamefont {Kumar}}, \bibinfo {author} {\bibfnamefont {L.}~\bibnamefont {Muechler}}, \bibinfo {author} {\bibfnamefont {A.}~\bibnamefont {Sun}}, \bibinfo {author} {\bibfnamefont {L.}~\bibnamefont {Jiao}}, \bibinfo {author} {\bibfnamefont {S.-Y.}\ \bibnamefont {Yang}}, \bibinfo {author} {\bibfnamefont {D.}~\bibnamefont {Liu}}, \bibinfo {author} {\bibfnamefont {A.}~\bibnamefont {Liang}}, \bibinfo {author} {\bibfnamefont {Q.}~\bibnamefont {Xu}}, \bibinfo {author} {\bibfnamefont {J.}~\bibnamefont {Kroder}}, \bibinfo {author} {\bibfnamefont {V.}~\bibnamefont {Süß}}, \bibinfo {author} {\bibfnamefont {H.}~\bibnamefont {Borrmann}}, \bibinfo {author} {\bibfnamefont {C.}~\bibnamefont {Shekhar}}, \bibinfo {author} {\bibfnamefont {Z.}~\bibnamefont {Wang}}, \bibinfo {author} {\bibfnamefont {C.}~\bibnamefont {Xi}}, \bibinfo {author}
  {\bibfnamefont {W.}~\bibnamefont {Wang}}, \bibinfo {author} {\bibfnamefont {W.}~\bibnamefont {Schnelle}}, \bibinfo {author} {\bibfnamefont {S.}~\bibnamefont {Wirth}}, \bibinfo {author} {\bibfnamefont {Y.}~\bibnamefont {Chen}}, \bibinfo {author} {\bibfnamefont {S.~T.~B.}\ \bibnamefont {Goennenwein}},\ and\ \bibinfo {author} {\bibfnamefont {C.}~\bibnamefont {Felser}},\ }\href {https://doi.org/10.1038/s41567-018-0234-5} {\bibfield  {journal} {\bibinfo  {journal} {Nature Physics}\ }\textbf {\bibinfo {volume} {14}},\ \bibinfo {pages} {1125} (\bibinfo {year} {2018})}\BibitemShut {NoStop}%
\bibitem [{\citenamefont {Liu}\ \emph {et~al.}(2022)\citenamefont {Liu}, \citenamefont {Liu}, \citenamefont {Xu}, \citenamefont {Shen}, \citenamefont {Li}, \citenamefont {Pei}, \citenamefont {Liang}, \citenamefont {Dudin}, \citenamefont {Kim}, \citenamefont {Cacho}, \citenamefont {Xu}, \citenamefont {Sun}, \citenamefont {Yang}, \citenamefont {Liu}, \citenamefont {Felser}, \citenamefont {Parkin},\ and\ \citenamefont {Chen}}]{Liu2022}%
  \BibitemOpen
  \bibfield  {author} {\bibinfo {author} {\bibfnamefont {D.~F.}\ \bibnamefont {Liu}}, \bibinfo {author} {\bibfnamefont {E.~K.}\ \bibnamefont {Liu}}, \bibinfo {author} {\bibfnamefont {Q.~N.}\ \bibnamefont {Xu}}, \bibinfo {author} {\bibfnamefont {J.~L.}\ \bibnamefont {Shen}}, \bibinfo {author} {\bibfnamefont {Y.~W.}\ \bibnamefont {Li}}, \bibinfo {author} {\bibfnamefont {D.}~\bibnamefont {Pei}}, \bibinfo {author} {\bibfnamefont {A.~J.}\ \bibnamefont {Liang}}, \bibinfo {author} {\bibfnamefont {P.}~\bibnamefont {Dudin}}, \bibinfo {author} {\bibfnamefont {T.~K.}\ \bibnamefont {Kim}}, \bibinfo {author} {\bibfnamefont {C.}~\bibnamefont {Cacho}}, \bibinfo {author} {\bibfnamefont {Y.~F.}\ \bibnamefont {Xu}}, \bibinfo {author} {\bibfnamefont {Y.}~\bibnamefont {Sun}}, \bibinfo {author} {\bibfnamefont {L.~X.}\ \bibnamefont {Yang}}, \bibinfo {author} {\bibfnamefont {Z.~K.}\ \bibnamefont {Liu}}, \bibinfo {author} {\bibfnamefont {C.}~\bibnamefont {Felser}}, \bibinfo {author} {\bibfnamefont {S.~S.~P.}\ \bibnamefont
  {Parkin}},\ and\ \bibinfo {author} {\bibfnamefont {Y.~L.}\ \bibnamefont {Chen}},\ }\href {https://doi.org/10.1038/s41535-021-00392-9} {\bibfield  {journal} {\bibinfo  {journal} {npj Quantum Materials}\ }\textbf {\bibinfo {volume} {7}},\ \bibinfo {pages} {11} (\bibinfo {year} {2022})}\BibitemShut {NoStop}%
\bibitem [{\citenamefont {Belopolski}\ \emph {et~al.}(2021)\citenamefont {Belopolski}, \citenamefont {Cochran}, \citenamefont {Liu}, \citenamefont {Cheng}, \citenamefont {Yang}, \citenamefont {Guguchia}, \citenamefont {Tsirkin}, \citenamefont {Yin}, \citenamefont {Vir}, \citenamefont {Thakur}, \citenamefont {Zhang}, \citenamefont {Zhang}, \citenamefont {Kaznatcheev}, \citenamefont {Cheng}, \citenamefont {Chang}, \citenamefont {Multer}, \citenamefont {Shumiya}, \citenamefont {Litskevich}, \citenamefont {Vescovo}, \citenamefont {Kim}, \citenamefont {Cacho}, \citenamefont {Yao}, \citenamefont {Felser}, \citenamefont {Neupert},\ and\ \citenamefont {Hasan}}]{Belopolski2021PRL}%
  \BibitemOpen
  \bibfield  {author} {\bibinfo {author} {\bibfnamefont {I.}~\bibnamefont {Belopolski}}, \bibinfo {author} {\bibfnamefont {T.~A.}\ \bibnamefont {Cochran}}, \bibinfo {author} {\bibfnamefont {X.}~\bibnamefont {Liu}}, \bibinfo {author} {\bibfnamefont {Z.-J.}\ \bibnamefont {Cheng}}, \bibinfo {author} {\bibfnamefont {X.~P.}\ \bibnamefont {Yang}}, \bibinfo {author} {\bibfnamefont {Z.}~\bibnamefont {Guguchia}}, \bibinfo {author} {\bibfnamefont {S.~S.}\ \bibnamefont {Tsirkin}}, \bibinfo {author} {\bibfnamefont {J.-X.}\ \bibnamefont {Yin}}, \bibinfo {author} {\bibfnamefont {P.}~\bibnamefont {Vir}}, \bibinfo {author} {\bibfnamefont {G.~S.}\ \bibnamefont {Thakur}}, \bibinfo {author} {\bibfnamefont {S.~S.}\ \bibnamefont {Zhang}}, \bibinfo {author} {\bibfnamefont {J.}~\bibnamefont {Zhang}}, \bibinfo {author} {\bibfnamefont {K.}~\bibnamefont {Kaznatcheev}}, \bibinfo {author} {\bibfnamefont {G.}~\bibnamefont {Cheng}}, \bibinfo {author} {\bibfnamefont {G.}~\bibnamefont {Chang}}, \bibinfo {author} {\bibfnamefont
  {D.}~\bibnamefont {Multer}}, \bibinfo {author} {\bibfnamefont {N.}~\bibnamefont {Shumiya}}, \bibinfo {author} {\bibfnamefont {M.}~\bibnamefont {Litskevich}}, \bibinfo {author} {\bibfnamefont {E.}~\bibnamefont {Vescovo}}, \bibinfo {author} {\bibfnamefont {T.~K.}\ \bibnamefont {Kim}}, \bibinfo {author} {\bibfnamefont {C.}~\bibnamefont {Cacho}}, \bibinfo {author} {\bibfnamefont {N.}~\bibnamefont {Yao}}, \bibinfo {author} {\bibfnamefont {C.}~\bibnamefont {Felser}}, \bibinfo {author} {\bibfnamefont {T.}~\bibnamefont {Neupert}},\ and\ \bibinfo {author} {\bibfnamefont {M.~Z.}\ \bibnamefont {Hasan}},\ }\href {https://doi.org/10.1103/PhysRevLett.127.256403} {\bibfield  {journal} {\bibinfo  {journal} {Physical Review Letters}\ }\textbf {\bibinfo {volume} {127}},\ \bibinfo {pages} {256403} (\bibinfo {year} {2021})}\BibitemShut {NoStop}%
\bibitem [{\citenamefont {Liu}\ \emph {et~al.}(2021)\citenamefont {Liu}, \citenamefont {Xu}, \citenamefont {Liu}, \citenamefont {Shen}, \citenamefont {Le}, \citenamefont {Li}, \citenamefont {Pei}, \citenamefont {Liang}, \citenamefont {Dudin}, \citenamefont {Kim}, \citenamefont {Cacho}, \citenamefont {Xu}, \citenamefont {Sun}, \citenamefont {Yang}, \citenamefont {Liu}, \citenamefont {Felser}, \citenamefont {Parkin},\ and\ \citenamefont {Chen}}]{DFLiu2021}%
  \BibitemOpen
  \bibfield  {author} {\bibinfo {author} {\bibfnamefont {D.~F.}\ \bibnamefont {Liu}}, \bibinfo {author} {\bibfnamefont {Q.~N.}\ \bibnamefont {Xu}}, \bibinfo {author} {\bibfnamefont {E.~K.}\ \bibnamefont {Liu}}, \bibinfo {author} {\bibfnamefont {J.~L.}\ \bibnamefont {Shen}}, \bibinfo {author} {\bibfnamefont {C.~C.}\ \bibnamefont {Le}}, \bibinfo {author} {\bibfnamefont {Y.~W.}\ \bibnamefont {Li}}, \bibinfo {author} {\bibfnamefont {D.}~\bibnamefont {Pei}}, \bibinfo {author} {\bibfnamefont {A.~J.}\ \bibnamefont {Liang}}, \bibinfo {author} {\bibfnamefont {P.}~\bibnamefont {Dudin}}, \bibinfo {author} {\bibfnamefont {T.~K.}\ \bibnamefont {Kim}}, \bibinfo {author} {\bibfnamefont {C.}~\bibnamefont {Cacho}}, \bibinfo {author} {\bibfnamefont {Y.~F.}\ \bibnamefont {Xu}}, \bibinfo {author} {\bibfnamefont {Y.}~\bibnamefont {Sun}}, \bibinfo {author} {\bibfnamefont {L.~X.}\ \bibnamefont {Yang}}, \bibinfo {author} {\bibfnamefont {Z.~K.}\ \bibnamefont {Liu}}, \bibinfo {author} {\bibfnamefont {C.}~\bibnamefont {Felser}},
  \bibinfo {author} {\bibfnamefont {S.~S.~P.}\ \bibnamefont {Parkin}},\ and\ \bibinfo {author} {\bibfnamefont {Y.~L.}\ \bibnamefont {Chen}},\ }\href {https://doi.org/10.1103/PhysRevB.104.205140} {\bibfield  {journal} {\bibinfo  {journal} {Physical Review B}\ }\textbf {\bibinfo {volume} {104}},\ \bibinfo {pages} {205140} (\bibinfo {year} {2021})}\BibitemShut {NoStop}%
\bibitem [{\citenamefont {Rossi}\ \emph {et~al.}(2021)\citenamefont {Rossi}, \citenamefont {Ivanov}, \citenamefont {Sreedhar}, \citenamefont {Gross}, \citenamefont {Shen}, \citenamefont {Rotenberg}, \citenamefont {Bostwick}, \citenamefont {Jozwiak}, \citenamefont {Taufour}, \citenamefont {Savrasov},\ and\ \citenamefont {Vishik}}]{Rossi2021}%
  \BibitemOpen
  \bibfield  {author} {\bibinfo {author} {\bibfnamefont {A.}~\bibnamefont {Rossi}}, \bibinfo {author} {\bibfnamefont {V.}~\bibnamefont {Ivanov}}, \bibinfo {author} {\bibfnamefont {S.}~\bibnamefont {Sreedhar}}, \bibinfo {author} {\bibfnamefont {A.~L.}\ \bibnamefont {Gross}}, \bibinfo {author} {\bibfnamefont {Z.}~\bibnamefont {Shen}}, \bibinfo {author} {\bibfnamefont {E.}~\bibnamefont {Rotenberg}}, \bibinfo {author} {\bibfnamefont {A.}~\bibnamefont {Bostwick}}, \bibinfo {author} {\bibfnamefont {C.}~\bibnamefont {Jozwiak}}, \bibinfo {author} {\bibfnamefont {V.}~\bibnamefont {Taufour}}, \bibinfo {author} {\bibfnamefont {S.~Y.}\ \bibnamefont {Savrasov}},\ and\ \bibinfo {author} {\bibfnamefont {I.~M.}\ \bibnamefont {Vishik}},\ }\href {https://doi.org/10.1103/PhysRevB.104.155115} {\bibfield  {journal} {\bibinfo  {journal} {Physical Review B}\ }\textbf {\bibinfo {volume} {104}},\ \bibinfo {pages} {155115} (\bibinfo {year} {2021})}\BibitemShut {NoStop}%
\bibitem [{\citenamefont {Ma}\ \emph {et~al.}(2019)\citenamefont {Ma}, \citenamefont {Nie}, \citenamefont {Yi}, \citenamefont {Jandke}, \citenamefont {Shang}, \citenamefont {Yao}, \citenamefont {Naamneh}, \citenamefont {Yan}, \citenamefont {Sun}, \citenamefont {Chikina}, \citenamefont {Strocov}, \citenamefont {Medarde}, \citenamefont {Song}, \citenamefont {Xiong}, \citenamefont {Xu}, \citenamefont {Wulfhekel}, \citenamefont {Mesot}, \citenamefont {Reticcioli}, \citenamefont {Franchini}, \citenamefont {Mudry}, \citenamefont {Müller}, \citenamefont {Shi}, \citenamefont {Qian}, \citenamefont {Ding},\ and\ \citenamefont {Shi}}]{Ma2019}%
  \BibitemOpen
  \bibfield  {author} {\bibinfo {author} {\bibfnamefont {J.-Z.}\ \bibnamefont {Ma}}, \bibinfo {author} {\bibfnamefont {S.~M.}\ \bibnamefont {Nie}}, \bibinfo {author} {\bibfnamefont {C.~J.}\ \bibnamefont {Yi}}, \bibinfo {author} {\bibfnamefont {J.}~\bibnamefont {Jandke}}, \bibinfo {author} {\bibfnamefont {T.}~\bibnamefont {Shang}}, \bibinfo {author} {\bibfnamefont {M.~Y.}\ \bibnamefont {Yao}}, \bibinfo {author} {\bibfnamefont {M.}~\bibnamefont {Naamneh}}, \bibinfo {author} {\bibfnamefont {L.~Q.}\ \bibnamefont {Yan}}, \bibinfo {author} {\bibfnamefont {Y.}~\bibnamefont {Sun}}, \bibinfo {author} {\bibfnamefont {A.}~\bibnamefont {Chikina}}, \bibinfo {author} {\bibfnamefont {V.~N.}\ \bibnamefont {Strocov}}, \bibinfo {author} {\bibfnamefont {M.}~\bibnamefont {Medarde}}, \bibinfo {author} {\bibfnamefont {M.}~\bibnamefont {Song}}, \bibinfo {author} {\bibfnamefont {Y.-M.}\ \bibnamefont {Xiong}}, \bibinfo {author} {\bibfnamefont {G.}~\bibnamefont {Xu}}, \bibinfo {author} {\bibfnamefont {W.}~\bibnamefont {Wulfhekel}},
  \bibinfo {author} {\bibfnamefont {J.}~\bibnamefont {Mesot}}, \bibinfo {author} {\bibfnamefont {M.}~\bibnamefont {Reticcioli}}, \bibinfo {author} {\bibfnamefont {C.}~\bibnamefont {Franchini}}, \bibinfo {author} {\bibfnamefont {C.}~\bibnamefont {Mudry}}, \bibinfo {author} {\bibfnamefont {M.}~\bibnamefont {Müller}}, \bibinfo {author} {\bibfnamefont {Y.~G.}\ \bibnamefont {Shi}}, \bibinfo {author} {\bibfnamefont {T.}~\bibnamefont {Qian}}, \bibinfo {author} {\bibfnamefont {H.}~\bibnamefont {Ding}},\ and\ \bibinfo {author} {\bibfnamefont {M.}~\bibnamefont {Shi}},\ }\href {https://doi.org/10.1126/sciadv.aaw4718} {\bibfield  {journal} {\bibinfo  {journal} {Science Advances}\ }\textbf {\bibinfo {volume} {5}},\ \bibinfo {pages} {eaaw4718} (\bibinfo {year} {2019})}\BibitemShut {NoStop}%
\bibitem [{\citenamefont {Sprague}\ \emph {et~al.}(2024)\citenamefont {Sprague}, \citenamefont {Regmi}, \citenamefont {Ghosh}, \citenamefont {Sakhya}, \citenamefont {Mondal}, \citenamefont {Elius}, \citenamefont {Valadez}, \citenamefont {Singh}, \citenamefont {Romanova}, \citenamefont {Kaczorowski}, \citenamefont {Bansil},\ and\ \citenamefont {Neupane}}]{Milo2024}%
  \BibitemOpen
  \bibfield  {author} {\bibinfo {author} {\bibfnamefont {M.~X.}\ \bibnamefont {Sprague}}, \bibinfo {author} {\bibfnamefont {S.}~\bibnamefont {Regmi}}, \bibinfo {author} {\bibfnamefont {B.}~\bibnamefont {Ghosh}}, \bibinfo {author} {\bibfnamefont {A.~P.}\ \bibnamefont {Sakhya}}, \bibinfo {author} {\bibfnamefont {M.~I.}\ \bibnamefont {Mondal}}, \bibinfo {author} {\bibfnamefont {I.~B.}\ \bibnamefont {Elius}}, \bibinfo {author} {\bibfnamefont {N.}~\bibnamefont {Valadez}}, \bibinfo {author} {\bibfnamefont {B.}~\bibnamefont {Singh}}, \bibinfo {author} {\bibfnamefont {T.}~\bibnamefont {Romanova}}, \bibinfo {author} {\bibfnamefont {D.}~\bibnamefont {Kaczorowski}}, \bibinfo {author} {\bibfnamefont {A.}~\bibnamefont {Bansil}},\ and\ \bibinfo {author} {\bibfnamefont {M.}~\bibnamefont {Neupane}},\ }\href {https://doi.org/10.1103/PhysRevB.110.045130} {\bibfield  {journal} {\bibinfo  {journal} {Physical Review B}\ }\textbf {\bibinfo {volume} {110}},\ \bibinfo {pages} {045130} (\bibinfo {year} {2024})}\BibitemShut {NoStop}%
\bibitem [{\citenamefont {Li}\ \emph {et~al.}(1995)\citenamefont {Li}, \citenamefont {Pearson}, \citenamefont {Bader}, \citenamefont {McIlroy}, \citenamefont {Waldfried},\ and\ \citenamefont {Dowben}}]{Li1995}%
  \BibitemOpen
  \bibfield  {author} {\bibinfo {author} {\bibfnamefont {D.}~\bibnamefont {Li}}, \bibinfo {author} {\bibfnamefont {J.}~\bibnamefont {Pearson}}, \bibinfo {author} {\bibfnamefont {S.~D.}\ \bibnamefont {Bader}}, \bibinfo {author} {\bibfnamefont {D.~N.}\ \bibnamefont {McIlroy}}, \bibinfo {author} {\bibfnamefont {C.}~\bibnamefont {Waldfried}},\ and\ \bibinfo {author} {\bibfnamefont {P.~A.}\ \bibnamefont {Dowben}},\ }\href {https://doi.org/10.1103/PhysRevB.51.13895} {\bibfield  {journal} {\bibinfo  {journal} {Physical Review B}\ }\textbf {\bibinfo {volume} {51}},\ \bibinfo {pages} {13895} (\bibinfo {year} {1995})}\BibitemShut {NoStop}%
\bibitem [{\citenamefont {Weschke}\ \emph {et~al.}(1996)\citenamefont {Weschke}, \citenamefont {Schüssler-Langeheine}, \citenamefont {Meier}, \citenamefont {Fedorov}, \citenamefont {Starke}, \citenamefont {Hübinger},\ and\ \citenamefont {Kaindl}}]{Weschke1996}%
  \BibitemOpen
  \bibfield  {author} {\bibinfo {author} {\bibfnamefont {E.}~\bibnamefont {Weschke}}, \bibinfo {author} {\bibfnamefont {C.}~\bibnamefont {Schüssler-Langeheine}}, \bibinfo {author} {\bibfnamefont {R.}~\bibnamefont {Meier}}, \bibinfo {author} {\bibfnamefont {A.~V.}\ \bibnamefont {Fedorov}}, \bibinfo {author} {\bibfnamefont {K.}~\bibnamefont {Starke}}, \bibinfo {author} {\bibfnamefont {F.}~\bibnamefont {Hübinger}},\ and\ \bibinfo {author} {\bibfnamefont {G.}~\bibnamefont {Kaindl}},\ }\href {https://doi.org/10.1103/PhysRevLett.77.3415} {\bibfield  {journal} {\bibinfo  {journal} {Physical Review Letters}\ }\textbf {\bibinfo {volume} {77}},\ \bibinfo {pages} {3415} (\bibinfo {year} {1996})}\BibitemShut {NoStop}%
\bibitem [{\citenamefont {Donath}\ \emph {et~al.}(1996)\citenamefont {Donath}, \citenamefont {Gubanka},\ and\ \citenamefont {Passek}}]{Donath1996}%
  \BibitemOpen
  \bibfield  {author} {\bibinfo {author} {\bibfnamefont {M.}~\bibnamefont {Donath}}, \bibinfo {author} {\bibfnamefont {B.}~\bibnamefont {Gubanka}},\ and\ \bibinfo {author} {\bibfnamefont {F.}~\bibnamefont {Passek}},\ }\href {https://doi.org/10.1103/PhysRevLett.77.5138} {\bibfield  {journal} {\bibinfo  {journal} {Physical Review Letters}\ }\textbf {\bibinfo {volume} {77}},\ \bibinfo {pages} {5138} (\bibinfo {year} {1996})}\BibitemShut {NoStop}%
\bibitem [{\citenamefont {Getzlaff}\ \emph {et~al.}(1998)\citenamefont {Getzlaff}, \citenamefont {Bode}, \citenamefont {Heinze}, \citenamefont {Pascal},\ and\ \citenamefont {Wiesendanger}}]{Getzlaff1998}%
  \BibitemOpen
  \bibfield  {author} {\bibinfo {author} {\bibfnamefont {M.}~\bibnamefont {Getzlaff}}, \bibinfo {author} {\bibfnamefont {M.}~\bibnamefont {Bode}}, \bibinfo {author} {\bibfnamefont {S.}~\bibnamefont {Heinze}}, \bibinfo {author} {\bibfnamefont {R.}~\bibnamefont {Pascal}},\ and\ \bibinfo {author} {\bibfnamefont {R.}~\bibnamefont {Wiesendanger}},\ }\href {https://doi.org/10.1016/S0304-8853(97)01140-2} {\bibfield  {journal} {\bibinfo  {journal} {Journal of Magnetism and Magnetic Materials}\ }\textbf {\bibinfo {volume} {184}},\ \bibinfo {pages} {155} (\bibinfo {year} {1998})}\BibitemShut {NoStop}%
\bibitem [{\citenamefont {Maiti}\ \emph {et~al.}(2002)\citenamefont {Maiti}, \citenamefont {Malagoli}, \citenamefont {Dallmeyer},\ and\ \citenamefont {Carbone}}]{Maiti2002}%
  \BibitemOpen
  \bibfield  {author} {\bibinfo {author} {\bibfnamefont {K.}~\bibnamefont {Maiti}}, \bibinfo {author} {\bibfnamefont {M.~C.}\ \bibnamefont {Malagoli}}, \bibinfo {author} {\bibfnamefont {A.}~\bibnamefont {Dallmeyer}},\ and\ \bibinfo {author} {\bibfnamefont {C.}~\bibnamefont {Carbone}},\ }\href {https://doi.org/10.1103/PhysRevLett.88.167205} {\bibfield  {journal} {\bibinfo  {journal} {Physical Review Letters}\ }\textbf {\bibinfo {volume} {88}},\ \bibinfo {pages} {167205} (\bibinfo {year} {2002})}\BibitemShut {NoStop}%
\bibitem [{\citenamefont {Fedorov}\ \emph {et~al.}(2002)\citenamefont {Fedorov}, \citenamefont {Valla}, \citenamefont {Liu}, \citenamefont {Johnson}, \citenamefont {Weinert},\ and\ \citenamefont {Allen}}]{Fedorov2002}%
  \BibitemOpen
  \bibfield  {author} {\bibinfo {author} {\bibfnamefont {A.~V.}\ \bibnamefont {Fedorov}}, \bibinfo {author} {\bibfnamefont {T.}~\bibnamefont {Valla}}, \bibinfo {author} {\bibfnamefont {F.}~\bibnamefont {Liu}}, \bibinfo {author} {\bibfnamefont {P.~D.}\ \bibnamefont {Johnson}}, \bibinfo {author} {\bibfnamefont {M.}~\bibnamefont {Weinert}},\ and\ \bibinfo {author} {\bibfnamefont {P.~B.}\ \bibnamefont {Allen}},\ }\href {https://doi.org/10.1103/PhysRevB.65.212409} {\bibfield  {journal} {\bibinfo  {journal} {Physical Review B}\ }\textbf {\bibinfo {volume} {65}},\ \bibinfo {pages} {212409} (\bibinfo {year} {2002})}\BibitemShut {NoStop}%
\bibitem [{\citenamefont {Qin}\ \emph {et~al.}(2022)\citenamefont {Qin}, \citenamefont {Chen}, \citenamefont {Du}, \citenamefont {Du}, \citenamefont {Zhang}, \citenamefont {Yin}, \citenamefont {Zhou}, \citenamefont {Xu}, \citenamefont {Gu}, \citenamefont {Zhang}, \citenamefont {Zhao}, \citenamefont {Li}, \citenamefont {Mo}, \citenamefont {Liu}, \citenamefont {Zhang}, \citenamefont {Guo}, \citenamefont {Tang}, \citenamefont {Chen},\ and\ \citenamefont {Yang}}]{Qin2022}%
  \BibitemOpen
  \bibfield  {author} {\bibinfo {author} {\bibfnamefont {N.}~\bibnamefont {Qin}}, \bibinfo {author} {\bibfnamefont {C.}~\bibnamefont {Chen}}, \bibinfo {author} {\bibfnamefont {S.}~\bibnamefont {Du}}, \bibinfo {author} {\bibfnamefont {X.}~\bibnamefont {Du}}, \bibinfo {author} {\bibfnamefont {X.}~\bibnamefont {Zhang}}, \bibinfo {author} {\bibfnamefont {Z.}~\bibnamefont {Yin}}, \bibinfo {author} {\bibfnamefont {J.}~\bibnamefont {Zhou}}, \bibinfo {author} {\bibfnamefont {R.}~\bibnamefont {Xu}}, \bibinfo {author} {\bibfnamefont {X.}~\bibnamefont {Gu}}, \bibinfo {author} {\bibfnamefont {Q.}~\bibnamefont {Zhang}}, \bibinfo {author} {\bibfnamefont {W.}~\bibnamefont {Zhao}}, \bibinfo {author} {\bibfnamefont {Y.}~\bibnamefont {Li}}, \bibinfo {author} {\bibfnamefont {S.-K.}\ \bibnamefont {Mo}}, \bibinfo {author} {\bibfnamefont {Z.}~\bibnamefont {Liu}}, \bibinfo {author} {\bibfnamefont {S.}~\bibnamefont {Zhang}}, \bibinfo {author} {\bibfnamefont {Y.}~\bibnamefont {Guo}}, \bibinfo {author} {\bibfnamefont {P.}~\bibnamefont
  {Tang}}, \bibinfo {author} {\bibfnamefont {Y.}~\bibnamefont {Chen}},\ and\ \bibinfo {author} {\bibfnamefont {L.}~\bibnamefont {Yang}},\ }\href {https://doi.org/10.1103/PhysRevB.106.035129} {\bibfield  {journal} {\bibinfo  {journal} {Physical Review B}\ }\textbf {\bibinfo {volume} {106}},\ \bibinfo {pages} {035129} (\bibinfo {year} {2022})}\BibitemShut {NoStop}%
\bibitem [{\citenamefont {Wu}\ \emph {et~al.}(2024)\citenamefont {Wu}, \citenamefont {Hu}, \citenamefont {Xie}, \citenamefont {Jang}, \citenamefont {Huang}, \citenamefont {Guo}, \citenamefont {Wu}, \citenamefont {Hu}, \citenamefont {Yue}, \citenamefont {Shi}, \citenamefont {Basak}, \citenamefont {Ren}, \citenamefont {Yilmaz}, \citenamefont {Vescovo}, \citenamefont {Jozwiak}, \citenamefont {Bostwick}, \citenamefont {Rotenberg}, \citenamefont {Fedorov}, \citenamefont {Denlinger}, \citenamefont {Klewe}, \citenamefont {Shafer}, \citenamefont {Lu}, \citenamefont {Hashimoto}, \citenamefont {Kono}, \citenamefont {Frano}, \citenamefont {Birgeneau}, \citenamefont {Xu}, \citenamefont {Zhu}, \citenamefont {Dai}, \citenamefont {Chu},\ and\ \citenamefont {Yi}}]{Wu2023}%
  \BibitemOpen
  \bibfield  {author} {\bibinfo {author} {\bibfnamefont {H.}~\bibnamefont {Wu}}, \bibinfo {author} {\bibfnamefont {C.}~\bibnamefont {Hu}}, \bibinfo {author} {\bibfnamefont {Y.}~\bibnamefont {Xie}}, \bibinfo {author} {\bibfnamefont {B.~G.}\ \bibnamefont {Jang}}, \bibinfo {author} {\bibfnamefont {J.}~\bibnamefont {Huang}}, \bibinfo {author} {\bibfnamefont {Y.}~\bibnamefont {Guo}}, \bibinfo {author} {\bibfnamefont {S.}~\bibnamefont {Wu}}, \bibinfo {author} {\bibfnamefont {C.}~\bibnamefont {Hu}}, \bibinfo {author} {\bibfnamefont {Z.}~\bibnamefont {Yue}}, \bibinfo {author} {\bibfnamefont {Y.}~\bibnamefont {Shi}}, \bibinfo {author} {\bibfnamefont {R.}~\bibnamefont {Basak}}, \bibinfo {author} {\bibfnamefont {Z.}~\bibnamefont {Ren}}, \bibinfo {author} {\bibfnamefont {T.}~\bibnamefont {Yilmaz}}, \bibinfo {author} {\bibfnamefont {E.}~\bibnamefont {Vescovo}}, \bibinfo {author} {\bibfnamefont {C.}~\bibnamefont {Jozwiak}}, \bibinfo {author} {\bibfnamefont {A.}~\bibnamefont {Bostwick}}, \bibinfo {author} {\bibfnamefont
  {E.}~\bibnamefont {Rotenberg}}, \bibinfo {author} {\bibfnamefont {A.}~\bibnamefont {Fedorov}}, \bibinfo {author} {\bibfnamefont {J.~D.}\ \bibnamefont {Denlinger}}, \bibinfo {author} {\bibfnamefont {C.}~\bibnamefont {Klewe}}, \bibinfo {author} {\bibfnamefont {P.}~\bibnamefont {Shafer}}, \bibinfo {author} {\bibfnamefont {D.}~\bibnamefont {Lu}}, \bibinfo {author} {\bibfnamefont {M.}~\bibnamefont {Hashimoto}}, \bibinfo {author} {\bibfnamefont {J.}~\bibnamefont {Kono}}, \bibinfo {author} {\bibfnamefont {A.}~\bibnamefont {Frano}}, \bibinfo {author} {\bibfnamefont {R.~J.}\ \bibnamefont {Birgeneau}}, \bibinfo {author} {\bibfnamefont {X.}~\bibnamefont {Xu}}, \bibinfo {author} {\bibfnamefont {J.-X.}\ \bibnamefont {Zhu}}, \bibinfo {author} {\bibfnamefont {P.}~\bibnamefont {Dai}}, \bibinfo {author} {\bibfnamefont {J.-H.}\ \bibnamefont {Chu}},\ and\ \bibinfo {author} {\bibfnamefont {M.}~\bibnamefont {Yi}},\ }\href {https://doi.org/10.1103/PhysRevB.109.104410} {\bibfield  {journal} {\bibinfo  {journal} {Physical Review
  B}\ }\textbf {\bibinfo {volume} {109}},\ \bibinfo {pages} {104410} (\bibinfo {year} {2024})}\BibitemShut {NoStop}%
\bibitem [{\citenamefont {Zhong}\ \emph {et~al.}(2023)\citenamefont {Zhong}, \citenamefont {Peng}, \citenamefont {Huang}, \citenamefont {Guan}, \citenamefont {Hwang}, \citenamefont {Hsu}, \citenamefont {Hu}, \citenamefont {Jia}, \citenamefont {Moritz}, \citenamefont {Lu}, \citenamefont {Lee}, \citenamefont {Jia}, \citenamefont {Devereaux}, \citenamefont {Mo},\ and\ \citenamefont {Shen}}]{Zhong2023}%
  \BibitemOpen
  \bibfield  {author} {\bibinfo {author} {\bibfnamefont {Y.}~\bibnamefont {Zhong}}, \bibinfo {author} {\bibfnamefont {C.}~\bibnamefont {Peng}}, \bibinfo {author} {\bibfnamefont {H.}~\bibnamefont {Huang}}, \bibinfo {author} {\bibfnamefont {D.}~\bibnamefont {Guan}}, \bibinfo {author} {\bibfnamefont {J.}~\bibnamefont {Hwang}}, \bibinfo {author} {\bibfnamefont {K.~H.}\ \bibnamefont {Hsu}}, \bibinfo {author} {\bibfnamefont {Y.}~\bibnamefont {Hu}}, \bibinfo {author} {\bibfnamefont {C.}~\bibnamefont {Jia}}, \bibinfo {author} {\bibfnamefont {B.}~\bibnamefont {Moritz}}, \bibinfo {author} {\bibfnamefont {D.}~\bibnamefont {Lu}}, \bibinfo {author} {\bibfnamefont {J.-S.}\ \bibnamefont {Lee}}, \bibinfo {author} {\bibfnamefont {J.-F.}\ \bibnamefont {Jia}}, \bibinfo {author} {\bibfnamefont {T.~P.}\ \bibnamefont {Devereaux}}, \bibinfo {author} {\bibfnamefont {S.-K.}\ \bibnamefont {Mo}},\ and\ \bibinfo {author} {\bibfnamefont {Z.-X.}\ \bibnamefont {Shen}},\ }\href {https://doi.org/10.1038/s41467-023-40997-1} {\bibfield
  {journal} {\bibinfo  {journal} {Nature Communications}\ }\textbf {\bibinfo {volume} {14}},\ \bibinfo {pages} {5340} (\bibinfo {year} {2023})}\BibitemShut {NoStop}%
\bibitem [{\citenamefont {Cheng}\ \emph {et~al.}(2023)\citenamefont {Cheng}, \citenamefont {Belopolski}, \citenamefont {Tien}, \citenamefont {Cochran}, \citenamefont {Yang}, \citenamefont {Ma}, \citenamefont {Yin}, \citenamefont {Chen}, \citenamefont {Zhang}, \citenamefont {Jozwiak}, \citenamefont {Bostwick}, \citenamefont {Rotenberg}, \citenamefont {Cheng}, \citenamefont {Hossain}, \citenamefont {Zhang}, \citenamefont {Litskevich}, \citenamefont {Jiang}, \citenamefont {Yao}, \citenamefont {Schroeter}, \citenamefont {Strocov}, \citenamefont {Lian}, \citenamefont {Felser}, \citenamefont {Chang}, \citenamefont {Jia}, \citenamefont {Chang},\ and\ \citenamefont {Hasan}}]{Cheng2023}%
  \BibitemOpen
  \bibfield  {author} {\bibinfo {author} {\bibfnamefont {Z.}~\bibnamefont {Cheng}}, \bibinfo {author} {\bibfnamefont {I.}~\bibnamefont {Belopolski}}, \bibinfo {author} {\bibfnamefont {H.}~\bibnamefont {Tien}}, \bibinfo {author} {\bibfnamefont {T.~A.}\ \bibnamefont {Cochran}}, \bibinfo {author} {\bibfnamefont {X.~P.}\ \bibnamefont {Yang}}, \bibinfo {author} {\bibfnamefont {W.}~\bibnamefont {Ma}}, \bibinfo {author} {\bibfnamefont {J.}~\bibnamefont {Yin}}, \bibinfo {author} {\bibfnamefont {D.}~\bibnamefont {Chen}}, \bibinfo {author} {\bibfnamefont {J.}~\bibnamefont {Zhang}}, \bibinfo {author} {\bibfnamefont {C.}~\bibnamefont {Jozwiak}}, \bibinfo {author} {\bibfnamefont {A.}~\bibnamefont {Bostwick}}, \bibinfo {author} {\bibfnamefont {E.}~\bibnamefont {Rotenberg}}, \bibinfo {author} {\bibfnamefont {G.}~\bibnamefont {Cheng}}, \bibinfo {author} {\bibfnamefont {M.~S.}\ \bibnamefont {Hossain}}, \bibinfo {author} {\bibfnamefont {Q.}~\bibnamefont {Zhang}}, \bibinfo {author} {\bibfnamefont {M.}~\bibnamefont
  {Litskevich}}, \bibinfo {author} {\bibfnamefont {Y.}~\bibnamefont {Jiang}}, \bibinfo {author} {\bibfnamefont {N.}~\bibnamefont {Yao}}, \bibinfo {author} {\bibfnamefont {N.~B.~M.}\ \bibnamefont {Schroeter}}, \bibinfo {author} {\bibfnamefont {V.~N.}\ \bibnamefont {Strocov}}, \bibinfo {author} {\bibfnamefont {B.}~\bibnamefont {Lian}}, \bibinfo {author} {\bibfnamefont {C.}~\bibnamefont {Felser}}, \bibinfo {author} {\bibfnamefont {G.}~\bibnamefont {Chang}}, \bibinfo {author} {\bibfnamefont {S.}~\bibnamefont {Jia}}, \bibinfo {author} {\bibfnamefont {T.}~\bibnamefont {Chang}},\ and\ \bibinfo {author} {\bibfnamefont {M.~Z.}\ \bibnamefont {Hasan}},\ }\href {https://doi.org/10.1002/adma.202205927} {\bibfield  {journal} {\bibinfo  {journal} {Advanced Materials}\ }\textbf {\bibinfo {volume} {35}},\ \bibinfo {pages} {2205927} (\bibinfo {year} {2023})}\BibitemShut {NoStop}%
\bibitem [{\citenamefont {Lou}\ \emph {et~al.}(2024)\citenamefont {Lou}, \citenamefont {Zhou}, \citenamefont {Song}, \citenamefont {Fedorov}, \citenamefont {Tu}, \citenamefont {Jiang}, \citenamefont {Wang}, \citenamefont {Li}, \citenamefont {Liu}, \citenamefont {Chen}, \citenamefont {Rader}, \citenamefont {Büchner}, \citenamefont {Sun}, \citenamefont {Weng}, \citenamefont {Lei},\ and\ \citenamefont {Wang}}]{Lou2024}%
  \BibitemOpen
  \bibfield  {author} {\bibinfo {author} {\bibfnamefont {R.}~\bibnamefont {Lou}}, \bibinfo {author} {\bibfnamefont {L.}~\bibnamefont {Zhou}}, \bibinfo {author} {\bibfnamefont {W.}~\bibnamefont {Song}}, \bibinfo {author} {\bibfnamefont {A.}~\bibnamefont {Fedorov}}, \bibinfo {author} {\bibfnamefont {Z.}~\bibnamefont {Tu}}, \bibinfo {author} {\bibfnamefont {B.}~\bibnamefont {Jiang}}, \bibinfo {author} {\bibfnamefont {Q.}~\bibnamefont {Wang}}, \bibinfo {author} {\bibfnamefont {M.}~\bibnamefont {Li}}, \bibinfo {author} {\bibfnamefont {Z.}~\bibnamefont {Liu}}, \bibinfo {author} {\bibfnamefont {X.}~\bibnamefont {Chen}}, \bibinfo {author} {\bibfnamefont {O.}~\bibnamefont {Rader}}, \bibinfo {author} {\bibfnamefont {B.}~\bibnamefont {Büchner}}, \bibinfo {author} {\bibfnamefont {Y.}~\bibnamefont {Sun}}, \bibinfo {author} {\bibfnamefont {H.}~\bibnamefont {Weng}}, \bibinfo {author} {\bibfnamefont {H.}~\bibnamefont {Lei}},\ and\ \bibinfo {author} {\bibfnamefont {S.}~\bibnamefont {Wang}},\ }\href
  {https://doi.org/10.1038/s41467-024-53343-w} {\bibfield  {journal} {\bibinfo  {journal} {Nature Communications}\ }\textbf {\bibinfo {volume} {15}},\ \bibinfo {pages} {9823} (\bibinfo {year} {2024})}\BibitemShut {NoStop}%
\bibitem [{\citenamefont {Damascelli}(2004)}]{Dama2004}%
  \BibitemOpen
  \bibfield  {author} {\bibinfo {author} {\bibfnamefont {A.}~\bibnamefont {Damascelli}},\ }\href {https://doi.org/10.1238/Physica.Topical.109a00061} {\bibfield  {journal} {\bibinfo  {journal} {Physica Scripta}\ }\textbf {\bibinfo {volume} {T109}},\ \bibinfo {pages} {61} (\bibinfo {year} {2004})}\BibitemShut {NoStop}%
\bibitem [{\citenamefont {Sánchez-Barriga}\ \emph {et~al.}(2016)\citenamefont {Sánchez-Barriga}, \citenamefont {Varykhalov}, \citenamefont {Springholz}, \citenamefont {Steiner}, \citenamefont {Kirchschlager}, \citenamefont {Bauer}, \citenamefont {Caha}, \citenamefont {Schierle}, \citenamefont {Weschke}, \citenamefont {Ünal}, \citenamefont {Valencia}, \citenamefont {Dunst}, \citenamefont {Braun}, \citenamefont {Ebert}, \citenamefont {Minár}, \citenamefont {Golias}, \citenamefont {Yashina}, \citenamefont {Ney}, \citenamefont {Holý},\ and\ \citenamefont {Rader}}]{Sanchez-Barriga2016}%
  \BibitemOpen
  \bibfield  {author} {\bibinfo {author} {\bibfnamefont {J.}~\bibnamefont {Sánchez-Barriga}}, \bibinfo {author} {\bibfnamefont {A.}~\bibnamefont {Varykhalov}}, \bibinfo {author} {\bibfnamefont {G.}~\bibnamefont {Springholz}}, \bibinfo {author} {\bibfnamefont {H.}~\bibnamefont {Steiner}}, \bibinfo {author} {\bibfnamefont {R.}~\bibnamefont {Kirchschlager}}, \bibinfo {author} {\bibfnamefont {G.}~\bibnamefont {Bauer}}, \bibinfo {author} {\bibfnamefont {O.}~\bibnamefont {Caha}}, \bibinfo {author} {\bibfnamefont {E.}~\bibnamefont {Schierle}}, \bibinfo {author} {\bibfnamefont {E.}~\bibnamefont {Weschke}}, \bibinfo {author} {\bibfnamefont {A.~A.}\ \bibnamefont {Ünal}}, \bibinfo {author} {\bibfnamefont {S.}~\bibnamefont {Valencia}}, \bibinfo {author} {\bibfnamefont {M.}~\bibnamefont {Dunst}}, \bibinfo {author} {\bibfnamefont {J.}~\bibnamefont {Braun}}, \bibinfo {author} {\bibfnamefont {H.}~\bibnamefont {Ebert}}, \bibinfo {author} {\bibfnamefont {J.}~\bibnamefont {Minár}}, \bibinfo {author} {\bibfnamefont
  {E.}~\bibnamefont {Golias}}, \bibinfo {author} {\bibfnamefont {L.~V.}\ \bibnamefont {Yashina}}, \bibinfo {author} {\bibfnamefont {A.}~\bibnamefont {Ney}}, \bibinfo {author} {\bibfnamefont {V.}~\bibnamefont {Holý}},\ and\ \bibinfo {author} {\bibfnamefont {O.}~\bibnamefont {Rader}},\ }\href {https://doi.org/10.1038/ncomms10559} {\bibfield  {journal} {\bibinfo  {journal} {Nature Communications}\ }\textbf {\bibinfo {volume} {7}},\ \bibinfo {pages} {10559} (\bibinfo {year} {2016})}\BibitemShut {NoStop}%
\bibitem [{\citenamefont {Shikin}\ \emph {et~al.}(2020)\citenamefont {Shikin}, \citenamefont {Estyunin}, \citenamefont {Klimovskikh}, \citenamefont {Filnov}, \citenamefont {Schwier}, \citenamefont {Kumar}, \citenamefont {Miyamoto}, \citenamefont {Okuda}, \citenamefont {Kimura}, \citenamefont {Kuroda}, \citenamefont {Yaji}, \citenamefont {Shin}, \citenamefont {Takeda}, \citenamefont {Saitoh}, \citenamefont {Aliev}, \citenamefont {Mamedov}, \citenamefont {Amiraslanov}, \citenamefont {Babanly}, \citenamefont {Otrokov}, \citenamefont {Eremeev},\ and\ \citenamefont {Chulkov}}]{Shikin2020}%
  \BibitemOpen
  \bibfield  {author} {\bibinfo {author} {\bibfnamefont {A.~M.}\ \bibnamefont {Shikin}}, \bibinfo {author} {\bibfnamefont {D.~A.}\ \bibnamefont {Estyunin}}, \bibinfo {author} {\bibfnamefont {I.~I.}\ \bibnamefont {Klimovskikh}}, \bibinfo {author} {\bibfnamefont {S.~O.}\ \bibnamefont {Filnov}}, \bibinfo {author} {\bibfnamefont {E.~F.}\ \bibnamefont {Schwier}}, \bibinfo {author} {\bibfnamefont {S.}~\bibnamefont {Kumar}}, \bibinfo {author} {\bibfnamefont {K.}~\bibnamefont {Miyamoto}}, \bibinfo {author} {\bibfnamefont {T.}~\bibnamefont {Okuda}}, \bibinfo {author} {\bibfnamefont {A.}~\bibnamefont {Kimura}}, \bibinfo {author} {\bibfnamefont {K.}~\bibnamefont {Kuroda}}, \bibinfo {author} {\bibfnamefont {K.}~\bibnamefont {Yaji}}, \bibinfo {author} {\bibfnamefont {S.}~\bibnamefont {Shin}}, \bibinfo {author} {\bibfnamefont {Y.}~\bibnamefont {Takeda}}, \bibinfo {author} {\bibfnamefont {Y.}~\bibnamefont {Saitoh}}, \bibinfo {author} {\bibfnamefont {Z.~S.}\ \bibnamefont {Aliev}}, \bibinfo {author} {\bibfnamefont {N.~T.}\
  \bibnamefont {Mamedov}}, \bibinfo {author} {\bibfnamefont {I.~R.}\ \bibnamefont {Amiraslanov}}, \bibinfo {author} {\bibfnamefont {M.~B.}\ \bibnamefont {Babanly}}, \bibinfo {author} {\bibfnamefont {M.~M.}\ \bibnamefont {Otrokov}}, \bibinfo {author} {\bibfnamefont {S.~V.}\ \bibnamefont {Eremeev}},\ and\ \bibinfo {author} {\bibfnamefont {E.~V.}\ \bibnamefont {Chulkov}},\ }\href {https://doi.org/10.1038/s41598-020-70089-9} {\bibfield  {journal} {\bibinfo  {journal} {Scientific Reports}\ }\textbf {\bibinfo {volume} {10}},\ \bibinfo {pages} {13226} (\bibinfo {year} {2020})}\BibitemShut {NoStop}%
\bibitem [{\citenamefont {Garnica}\ \emph {et~al.}(2022)\citenamefont {Garnica}, \citenamefont {Otrokov}, \citenamefont {Aguilar}, \citenamefont {Klimovskikh}, \citenamefont {Estyunin}, \citenamefont {Aliev}, \citenamefont {Amiraslanov}, \citenamefont {Abdullayev}, \citenamefont {Zverev}, \citenamefont {Babanly}, \citenamefont {Mamedov}, \citenamefont {Shikin}, \citenamefont {Arnau}, \citenamefont {de~Parga}, \citenamefont {Chulkov},\ and\ \citenamefont {Miranda}}]{Garnica2022}%
  \BibitemOpen
  \bibfield  {author} {\bibinfo {author} {\bibfnamefont {M.}~\bibnamefont {Garnica}}, \bibinfo {author} {\bibfnamefont {M.~M.}\ \bibnamefont {Otrokov}}, \bibinfo {author} {\bibfnamefont {P.~C.}\ \bibnamefont {Aguilar}}, \bibinfo {author} {\bibfnamefont {I.~I.}\ \bibnamefont {Klimovskikh}}, \bibinfo {author} {\bibfnamefont {D.}~\bibnamefont {Estyunin}}, \bibinfo {author} {\bibfnamefont {Z.~S.}\ \bibnamefont {Aliev}}, \bibinfo {author} {\bibfnamefont {I.~R.}\ \bibnamefont {Amiraslanov}}, \bibinfo {author} {\bibfnamefont {N.~A.}\ \bibnamefont {Abdullayev}}, \bibinfo {author} {\bibfnamefont {V.~N.}\ \bibnamefont {Zverev}}, \bibinfo {author} {\bibfnamefont {M.~B.}\ \bibnamefont {Babanly}}, \bibinfo {author} {\bibfnamefont {N.~T.}\ \bibnamefont {Mamedov}}, \bibinfo {author} {\bibfnamefont {A.~M.}\ \bibnamefont {Shikin}}, \bibinfo {author} {\bibfnamefont {A.}~\bibnamefont {Arnau}}, \bibinfo {author} {\bibfnamefont {A.~L.~V.}\ \bibnamefont {de~Parga}}, \bibinfo {author} {\bibfnamefont {E.~V.}\ \bibnamefont
  {Chulkov}},\ and\ \bibinfo {author} {\bibfnamefont {R.}~\bibnamefont {Miranda}},\ }\href {https://doi.org/10.1038/s41535-021-00414-6} {\bibfield  {journal} {\bibinfo  {journal} {npj Quantum Materials}\ }\textbf {\bibinfo {volume} {7}},\ \bibinfo {pages} {7} (\bibinfo {year} {2022})}\BibitemShut {NoStop}%
\bibitem [{\citenamefont {Chuang}\ \emph {et~al.}(2016)\citenamefont {Chuang}, \citenamefont {Hsu}, \citenamefont {Chou}, \citenamefont {Crisostomo}, \citenamefont {Huang}, \citenamefont {Wu}, \citenamefont {Kuo}, \citenamefont {Yeh}, \citenamefont {Lin},\ and\ \citenamefont {Bansil}}]{Chuang2016}%
  \BibitemOpen
  \bibfield  {author} {\bibinfo {author} {\bibfnamefont {F.-C.}\ \bibnamefont {Chuang}}, \bibinfo {author} {\bibfnamefont {C.-H.}\ \bibnamefont {Hsu}}, \bibinfo {author} {\bibfnamefont {H.-L.}\ \bibnamefont {Chou}}, \bibinfo {author} {\bibfnamefont {C.~P.}\ \bibnamefont {Crisostomo}}, \bibinfo {author} {\bibfnamefont {Z.-Q.}\ \bibnamefont {Huang}}, \bibinfo {author} {\bibfnamefont {S.-Y.}\ \bibnamefont {Wu}}, \bibinfo {author} {\bibfnamefont {C.-C.}\ \bibnamefont {Kuo}}, \bibinfo {author} {\bibfnamefont {W.-C.~V.}\ \bibnamefont {Yeh}}, \bibinfo {author} {\bibfnamefont {H.}~\bibnamefont {Lin}},\ and\ \bibinfo {author} {\bibfnamefont {A.}~\bibnamefont {Bansil}},\ }\href {https://doi.org/10.1103/PhysRevB.93.035429} {\bibfield  {journal} {\bibinfo  {journal} {Physical Review B}\ }\textbf {\bibinfo {volume} {93}},\ \bibinfo {pages} {035429} (\bibinfo {year} {2016})}\BibitemShut {NoStop}%
\bibitem [{\citenamefont {Feng}\ \emph {et~al.}(2017)\citenamefont {Feng}, \citenamefont {Fu}, \citenamefont {Kasamatsu}, \citenamefont {Ito}, \citenamefont {Cheng}, \citenamefont {Liu}, \citenamefont {Feng}, \citenamefont {Wu}, \citenamefont {Mahatha}, \citenamefont {Sheverdyaeva}, \citenamefont {Moras}, \citenamefont {Arita}, \citenamefont {Sugino}, \citenamefont {Chiang}, \citenamefont {Shimada}, \citenamefont {Miyamoto}, \citenamefont {Okuda}, \citenamefont {Wu}, \citenamefont {Chen}, \citenamefont {Yao},\ and\ \citenamefont {Matsuda}}]{Feng2017c}%
  \BibitemOpen
  \bibfield  {author} {\bibinfo {author} {\bibfnamefont {B.}~\bibnamefont {Feng}}, \bibinfo {author} {\bibfnamefont {B.}~\bibnamefont {Fu}}, \bibinfo {author} {\bibfnamefont {S.}~\bibnamefont {Kasamatsu}}, \bibinfo {author} {\bibfnamefont {S.}~\bibnamefont {Ito}}, \bibinfo {author} {\bibfnamefont {P.}~\bibnamefont {Cheng}}, \bibinfo {author} {\bibfnamefont {C.-C.}\ \bibnamefont {Liu}}, \bibinfo {author} {\bibfnamefont {Y.}~\bibnamefont {Feng}}, \bibinfo {author} {\bibfnamefont {S.}~\bibnamefont {Wu}}, \bibinfo {author} {\bibfnamefont {S.~K.}\ \bibnamefont {Mahatha}}, \bibinfo {author} {\bibfnamefont {P.}~\bibnamefont {Sheverdyaeva}}, \bibinfo {author} {\bibfnamefont {P.}~\bibnamefont {Moras}}, \bibinfo {author} {\bibfnamefont {M.}~\bibnamefont {Arita}}, \bibinfo {author} {\bibfnamefont {O.}~\bibnamefont {Sugino}}, \bibinfo {author} {\bibfnamefont {T.-C.}\ \bibnamefont {Chiang}}, \bibinfo {author} {\bibfnamefont {K.}~\bibnamefont {Shimada}}, \bibinfo {author} {\bibfnamefont {K.}~\bibnamefont {Miyamoto}},
  \bibinfo {author} {\bibfnamefont {T.}~\bibnamefont {Okuda}}, \bibinfo {author} {\bibfnamefont {K.}~\bibnamefont {Wu}}, \bibinfo {author} {\bibfnamefont {L.}~\bibnamefont {Chen}}, \bibinfo {author} {\bibfnamefont {Y.}~\bibnamefont {Yao}},\ and\ \bibinfo {author} {\bibfnamefont {I.}~\bibnamefont {Matsuda}},\ }\href {https://doi.org/10.1038/s41467-017-01108-z} {\bibfield  {journal} {\bibinfo  {journal} {Nature Communications}\ }\textbf {\bibinfo {volume} {8}},\ \bibinfo {pages} {1007} (\bibinfo {year} {2017})}\BibitemShut {NoStop}%
\bibitem [{\citenamefont {Feng}\ \emph {et~al.}(2019)\citenamefont {Feng}, \citenamefont {Zhang}, \citenamefont {Feng}, \citenamefont {Fu}, \citenamefont {Wu}, \citenamefont {Miyamoto}, \citenamefont {He}, \citenamefont {Chen}, \citenamefont {Wu}, \citenamefont {Shimada}, \citenamefont {Okuda},\ and\ \citenamefont {Yao}}]{Feng2019a}%
  \BibitemOpen
  \bibfield  {author} {\bibinfo {author} {\bibfnamefont {B.}~\bibnamefont {Feng}}, \bibinfo {author} {\bibfnamefont {R.-W.}\ \bibnamefont {Zhang}}, \bibinfo {author} {\bibfnamefont {Y.}~\bibnamefont {Feng}}, \bibinfo {author} {\bibfnamefont {B.}~\bibnamefont {Fu}}, \bibinfo {author} {\bibfnamefont {S.}~\bibnamefont {Wu}}, \bibinfo {author} {\bibfnamefont {K.}~\bibnamefont {Miyamoto}}, \bibinfo {author} {\bibfnamefont {S.}~\bibnamefont {He}}, \bibinfo {author} {\bibfnamefont {L.}~\bibnamefont {Chen}}, \bibinfo {author} {\bibfnamefont {K.}~\bibnamefont {Wu}}, \bibinfo {author} {\bibfnamefont {K.}~\bibnamefont {Shimada}}, \bibinfo {author} {\bibfnamefont {T.}~\bibnamefont {Okuda}},\ and\ \bibinfo {author} {\bibfnamefont {Y.}~\bibnamefont {Yao}},\ }\href {https://doi.org/10.1103/PhysRevLett.123.116401} {\bibfield  {journal} {\bibinfo  {journal} {Physical Review Letters}\ }\textbf {\bibinfo {volume} {123}},\ \bibinfo {pages} {116401} (\bibinfo {year} {2019})}\BibitemShut {NoStop}%
\bibitem [{\citenamefont {Ünzelmann}\ \emph {et~al.}(2020)\citenamefont {Ünzelmann}, \citenamefont {Bentmann}, \citenamefont {Eck}, \citenamefont {Kißlinger}, \citenamefont {Geldiyev}, \citenamefont {Rieger}, \citenamefont {Moser}, \citenamefont {Vidal}, \citenamefont {Kißner}, \citenamefont {Hammer}, \citenamefont {Schneider}, \citenamefont {Fauster}, \citenamefont {Sangiovanni}, \citenamefont {Sante},\ and\ \citenamefont {Reinert}}]{Unzelmann2020a}%
  \BibitemOpen
  \bibfield  {author} {\bibinfo {author} {\bibfnamefont {M.}~\bibnamefont {Ünzelmann}}, \bibinfo {author} {\bibfnamefont {H.}~\bibnamefont {Bentmann}}, \bibinfo {author} {\bibfnamefont {P.}~\bibnamefont {Eck}}, \bibinfo {author} {\bibfnamefont {T.}~\bibnamefont {Kißlinger}}, \bibinfo {author} {\bibfnamefont {B.}~\bibnamefont {Geldiyev}}, \bibinfo {author} {\bibfnamefont {J.}~\bibnamefont {Rieger}}, \bibinfo {author} {\bibfnamefont {S.}~\bibnamefont {Moser}}, \bibinfo {author} {\bibfnamefont {R.~C.}\ \bibnamefont {Vidal}}, \bibinfo {author} {\bibfnamefont {K.}~\bibnamefont {Kißner}}, \bibinfo {author} {\bibfnamefont {L.}~\bibnamefont {Hammer}}, \bibinfo {author} {\bibfnamefont {M.~A.}\ \bibnamefont {Schneider}}, \bibinfo {author} {\bibfnamefont {T.}~\bibnamefont {Fauster}}, \bibinfo {author} {\bibfnamefont {G.}~\bibnamefont {Sangiovanni}}, \bibinfo {author} {\bibfnamefont {D.~D.}\ \bibnamefont {Sante}},\ and\ \bibinfo {author} {\bibfnamefont {F.}~\bibnamefont {Reinert}},\ }\href
  {https://doi.org/10.1103/PhysRevLett.124.176401} {\bibfield  {journal} {\bibinfo  {journal} {Physical Review Letters}\ }\textbf {\bibinfo {volume} {124}},\ \bibinfo {pages} {176401} (\bibinfo {year} {2020})}\BibitemShut {NoStop}%
\bibitem [{\citenamefont {Liu}\ \emph {et~al.}(2020)\citenamefont {Liu}, \citenamefont {Wang}, \citenamefont {Li}, \citenamefont {Chen}, \citenamefont {Jia},\ and\ \citenamefont {Cho}}]{Liu2020}%
  \BibitemOpen
  \bibfield  {author} {\bibinfo {author} {\bibfnamefont {L.}~\bibnamefont {Liu}}, \bibinfo {author} {\bibfnamefont {C.}~\bibnamefont {Wang}}, \bibinfo {author} {\bibfnamefont {J.}~\bibnamefont {Li}}, \bibinfo {author} {\bibfnamefont {X.~Q.}\ \bibnamefont {Chen}}, \bibinfo {author} {\bibfnamefont {Y.}~\bibnamefont {Jia}},\ and\ \bibinfo {author} {\bibfnamefont {J.~H.}\ \bibnamefont {Cho}},\ }\href {https://doi.org/10.1103/PhysRevB.101.165403} {\bibfield  {journal} {\bibinfo  {journal} {Physical Review B}\ }\textbf {\bibinfo {volume} {101}},\ \bibinfo {pages} {1} (\bibinfo {year} {2020})}\BibitemShut {NoStop}%
\bibitem [{\citenamefont {Lu}\ \emph {et~al.}(2021)\citenamefont {Lu}, \citenamefont {Gao}, \citenamefont {Song}, \citenamefont {Li}, \citenamefont {Niu}, \citenamefont {Chen}, \citenamefont {Qian}, \citenamefont {Ding}, \citenamefont {Lin}, \citenamefont {Du},\ and\ \citenamefont {Gao}}]{Lu2021}%
  \BibitemOpen
  \bibfield  {author} {\bibinfo {author} {\bibfnamefont {J.}~\bibnamefont {Lu}}, \bibinfo {author} {\bibfnamefont {L.}~\bibnamefont {Gao}}, \bibinfo {author} {\bibfnamefont {S.}~\bibnamefont {Song}}, \bibinfo {author} {\bibfnamefont {H.}~\bibnamefont {Li}}, \bibinfo {author} {\bibfnamefont {G.}~\bibnamefont {Niu}}, \bibinfo {author} {\bibfnamefont {H.}~\bibnamefont {Chen}}, \bibinfo {author} {\bibfnamefont {T.}~\bibnamefont {Qian}}, \bibinfo {author} {\bibfnamefont {H.}~\bibnamefont {Ding}}, \bibinfo {author} {\bibfnamefont {X.}~\bibnamefont {Lin}}, \bibinfo {author} {\bibfnamefont {S.}~\bibnamefont {Du}},\ and\ \bibinfo {author} {\bibfnamefont {H.~J.}\ \bibnamefont {Gao}},\ }\href {https://doi.org/10.1021/acsanm.1c01517} {\bibfield  {journal} {\bibinfo  {journal} {ACS Applied Nano Materials}\ }\textbf {\bibinfo {volume} {4}},\ \bibinfo {pages} {8845} (\bibinfo {year} {2021})}\BibitemShut {NoStop}%
\bibitem [{\citenamefont {Bauernfeind}\ \emph {et~al.}(2021)\citenamefont {Bauernfeind}, \citenamefont {Erhardt}, \citenamefont {Eck}, \citenamefont {Thakur}, \citenamefont {Gabel}, \citenamefont {Lee}, \citenamefont {Schäfer}, \citenamefont {Moser}, \citenamefont {Sante}, \citenamefont {Claessen},\ and\ \citenamefont {Sangiovanni}}]{Bauernfeind2021}%
  \BibitemOpen
  \bibfield  {author} {\bibinfo {author} {\bibfnamefont {M.}~\bibnamefont {Bauernfeind}}, \bibinfo {author} {\bibfnamefont {J.}~\bibnamefont {Erhardt}}, \bibinfo {author} {\bibfnamefont {P.}~\bibnamefont {Eck}}, \bibinfo {author} {\bibfnamefont {P.~K.}\ \bibnamefont {Thakur}}, \bibinfo {author} {\bibfnamefont {J.}~\bibnamefont {Gabel}}, \bibinfo {author} {\bibfnamefont {T.-L.}\ \bibnamefont {Lee}}, \bibinfo {author} {\bibfnamefont {J.}~\bibnamefont {Schäfer}}, \bibinfo {author} {\bibfnamefont {S.}~\bibnamefont {Moser}}, \bibinfo {author} {\bibfnamefont {D.~D.}\ \bibnamefont {Sante}}, \bibinfo {author} {\bibfnamefont {R.}~\bibnamefont {Claessen}},\ and\ \bibinfo {author} {\bibfnamefont {G.}~\bibnamefont {Sangiovanni}},\ }\href {https://doi.org/10.1038/s41467-021-25627-y} {\bibfield  {journal} {\bibinfo  {journal} {Nature Communications}\ }\textbf {\bibinfo {volume} {12}},\ \bibinfo {pages} {5396} (\bibinfo {year} {2021})}\BibitemShut {NoStop}%
\bibitem [{\citenamefont {Cameau}\ \emph {et~al.}(2024)\citenamefont {Cameau}, \citenamefont {Olszowska}, \citenamefont {Rosmus}, \citenamefont {Silly}, \citenamefont {Cren}, \citenamefont {Malecot}, \citenamefont {David},\ and\ \citenamefont {D’angelo}}]{Cameau2024}%
  \BibitemOpen
  \bibfield  {author} {\bibinfo {author} {\bibfnamefont {M.}~\bibnamefont {Cameau}}, \bibinfo {author} {\bibfnamefont {N.}~\bibnamefont {Olszowska}}, \bibinfo {author} {\bibfnamefont {M.}~\bibnamefont {Rosmus}}, \bibinfo {author} {\bibfnamefont {M.~G.}\ \bibnamefont {Silly}}, \bibinfo {author} {\bibfnamefont {T.}~\bibnamefont {Cren}}, \bibinfo {author} {\bibfnamefont {A.}~\bibnamefont {Malecot}}, \bibinfo {author} {\bibfnamefont {P.}~\bibnamefont {David}},\ and\ \bibinfo {author} {\bibfnamefont {M.}~\bibnamefont {D’angelo}},\ }\href {https://doi.org/10.1088/2053-1583/ad471e} {\bibfield  {journal} {\bibinfo  {journal} {2D Materials}\ }\textbf {\bibinfo {volume} {11}},\ \bibinfo {pages} {035023} (\bibinfo {year} {2024})}\BibitemShut {NoStop}%
\bibitem [{\citenamefont {Matetskiy}\ \emph {et~al.}(2025)\citenamefont {Matetskiy}, \citenamefont {Barla}, \citenamefont {Moras}, \citenamefont {Carbone}, \citenamefont {Milotti}, \citenamefont {Brondin}, \citenamefont {Benher}, \citenamefont {Holub}, \citenamefont {Ohresser}, \citenamefont {Otero}, \citenamefont {Choueikani}, \citenamefont {Shvets}, \citenamefont {Mihalyuk}, \citenamefont {Eremeev},\ and\ \citenamefont {Sheverdyaeva}}]{Matetskiy2025}%
  \BibitemOpen
  \bibfield  {author} {\bibinfo {author} {\bibfnamefont {A.~V.}\ \bibnamefont {Matetskiy}}, \bibinfo {author} {\bibfnamefont {A.}~\bibnamefont {Barla}}, \bibinfo {author} {\bibfnamefont {P.}~\bibnamefont {Moras}}, \bibinfo {author} {\bibfnamefont {C.}~\bibnamefont {Carbone}}, \bibinfo {author} {\bibfnamefont {V.}~\bibnamefont {Milotti}}, \bibinfo {author} {\bibfnamefont {C.~A.}\ \bibnamefont {Brondin}}, \bibinfo {author} {\bibfnamefont {Z.~R.}\ \bibnamefont {Benher}}, \bibinfo {author} {\bibfnamefont {M.}~\bibnamefont {Holub}}, \bibinfo {author} {\bibfnamefont {P.}~\bibnamefont {Ohresser}}, \bibinfo {author} {\bibfnamefont {E.}~\bibnamefont {Otero}}, \bibinfo {author} {\bibfnamefont {F.}~\bibnamefont {Choueikani}}, \bibinfo {author} {\bibfnamefont {I.~A.}\ \bibnamefont {Shvets}}, \bibinfo {author} {\bibfnamefont {A.~N.}\ \bibnamefont {Mihalyuk}}, \bibinfo {author} {\bibfnamefont {S.~V.}\ \bibnamefont {Eremeev}},\ and\ \bibinfo {author} {\bibfnamefont {P.~M.}\ \bibnamefont {Sheverdyaeva}},\ }\href
  {https://doi.org/10.1021/acsnano.5c03331} {\bibfield  {journal} {\bibinfo  {journal} {ACS Nano}\ }\textbf {\bibinfo {volume} {19}},\ \bibinfo {pages} {20863} (\bibinfo {year} {2025})}\BibitemShut {NoStop}%
\bibitem [{\citenamefont {Ormaza}\ \emph {et~al.}(2016)\citenamefont {Ormaza}, \citenamefont {Fernández}, \citenamefont {Ilyn}, \citenamefont {Magaña}, \citenamefont {Xu}, \citenamefont {Verstraete}, \citenamefont {Gastaldo}, \citenamefont {Valbuena}, \citenamefont {Gargiani}, \citenamefont {Mugarza}, \citenamefont {Ayuela}, \citenamefont {Vitali}, \citenamefont {Blanco-Rey}, \citenamefont {Schiller},\ and\ \citenamefont {Ortega}}]{Ormaza2016}%
  \BibitemOpen
  \bibfield  {author} {\bibinfo {author} {\bibfnamefont {M.}~\bibnamefont {Ormaza}}, \bibinfo {author} {\bibfnamefont {L.}~\bibnamefont {Fernández}}, \bibinfo {author} {\bibfnamefont {M.}~\bibnamefont {Ilyn}}, \bibinfo {author} {\bibfnamefont {A.}~\bibnamefont {Magaña}}, \bibinfo {author} {\bibfnamefont {B.}~\bibnamefont {Xu}}, \bibinfo {author} {\bibfnamefont {M.~J.}\ \bibnamefont {Verstraete}}, \bibinfo {author} {\bibfnamefont {M.}~\bibnamefont {Gastaldo}}, \bibinfo {author} {\bibfnamefont {M.~A.}\ \bibnamefont {Valbuena}}, \bibinfo {author} {\bibfnamefont {P.}~\bibnamefont {Gargiani}}, \bibinfo {author} {\bibfnamefont {A.}~\bibnamefont {Mugarza}}, \bibinfo {author} {\bibfnamefont {A.}~\bibnamefont {Ayuela}}, \bibinfo {author} {\bibfnamefont {L.}~\bibnamefont {Vitali}}, \bibinfo {author} {\bibfnamefont {M.}~\bibnamefont {Blanco-Rey}}, \bibinfo {author} {\bibfnamefont {F.}~\bibnamefont {Schiller}},\ and\ \bibinfo {author} {\bibfnamefont {J.~E.}\ \bibnamefont {Ortega}},\ }\href
  {https://doi.org/10.1021/acs.nanolett.6b01197} {\bibfield  {journal} {\bibinfo  {journal} {Nano Letters}\ }\textbf {\bibinfo {volume} {16}},\ \bibinfo {pages} {4230} (\bibinfo {year} {2016})}\BibitemShut {NoStop}%
\bibitem [{sup()}]{supple1}%
  \BibitemOpen
  \href@noop {} {\bibinfo {title} {See supplementary information at [url] for the details of the methods and analysis.}}\BibitemShut {Stop}%
\bibitem [{\citenamefont {Correa}\ \emph {et~al.}(2016)\citenamefont {Correa}, \citenamefont {Xu}, \citenamefont {Verstraete},\ and\ \citenamefont {Vitali}}]{Correa2016}%
  \BibitemOpen
  \bibfield  {author} {\bibinfo {author} {\bibfnamefont {A.}~\bibnamefont {Correa}}, \bibinfo {author} {\bibfnamefont {B.}~\bibnamefont {Xu}}, \bibinfo {author} {\bibfnamefont {M.~J.}\ \bibnamefont {Verstraete}},\ and\ \bibinfo {author} {\bibfnamefont {L.}~\bibnamefont {Vitali}},\ }\href {https://doi.org/10.1039/C6NR06398E} {\bibfield  {journal} {\bibinfo  {journal} {Nanoscale}\ }\textbf {\bibinfo {volume} {8}},\ \bibinfo {pages} {19148} (\bibinfo {year} {2016})}\BibitemShut {NoStop}%
\bibitem [{\citenamefont {Fernandez}\ \emph {et~al.}(2020)\citenamefont {Fernandez}, \citenamefont {Blanco-Rey}, \citenamefont {Castrillo-Bodero}, \citenamefont {Ilyn}, \citenamefont {Ali}, \citenamefont {Turco}, \citenamefont {Corso}, \citenamefont {Ormaza}, \citenamefont {Gargiani}, \citenamefont {Valbuena}, \citenamefont {Mugarza}, \citenamefont {Moras}, \citenamefont {Sheverdyaeva}, \citenamefont {Kundu}, \citenamefont {Jugovac}, \citenamefont {Laubschat}, \citenamefont {Ortega},\ and\ \citenamefont {Schiller}}]{Fernandez2020}%
  \BibitemOpen
  \bibfield  {author} {\bibinfo {author} {\bibfnamefont {L.}~\bibnamefont {Fernandez}}, \bibinfo {author} {\bibfnamefont {M.}~\bibnamefont {Blanco-Rey}}, \bibinfo {author} {\bibfnamefont {R.}~\bibnamefont {Castrillo-Bodero}}, \bibinfo {author} {\bibfnamefont {M.}~\bibnamefont {Ilyn}}, \bibinfo {author} {\bibfnamefont {K.}~\bibnamefont {Ali}}, \bibinfo {author} {\bibfnamefont {E.}~\bibnamefont {Turco}}, \bibinfo {author} {\bibfnamefont {M.}~\bibnamefont {Corso}}, \bibinfo {author} {\bibfnamefont {M.}~\bibnamefont {Ormaza}}, \bibinfo {author} {\bibfnamefont {P.}~\bibnamefont {Gargiani}}, \bibinfo {author} {\bibfnamefont {M.~A.}\ \bibnamefont {Valbuena}}, \bibinfo {author} {\bibfnamefont {A.}~\bibnamefont {Mugarza}}, \bibinfo {author} {\bibfnamefont {P.}~\bibnamefont {Moras}}, \bibinfo {author} {\bibfnamefont {P.~M.}\ \bibnamefont {Sheverdyaeva}}, \bibinfo {author} {\bibfnamefont {A.~K.}\ \bibnamefont {Kundu}}, \bibinfo {author} {\bibfnamefont {M.}~\bibnamefont {Jugovac}}, \bibinfo {author} {\bibfnamefont
  {C.}~\bibnamefont {Laubschat}}, \bibinfo {author} {\bibfnamefont {J.~E.}\ \bibnamefont {Ortega}},\ and\ \bibinfo {author} {\bibfnamefont {F.}~\bibnamefont {Schiller}},\ }\href {https://doi.org/10.1039/D0NR04964F} {\bibfield  {journal} {\bibinfo  {journal} {Nanoscale}\ }\textbf {\bibinfo {volume} {12}},\ \bibinfo {pages} {22258} (\bibinfo {year} {2020})}\BibitemShut {NoStop}%
\bibitem [{\citenamefont {Bentmann}\ \emph {et~al.}(2012)\citenamefont {Bentmann}, \citenamefont {Abdelouahed}, \citenamefont {Mulazzi}, \citenamefont {Henk},\ and\ \citenamefont {Reinert}}]{Bentmann2012}%
  \BibitemOpen
  \bibfield  {author} {\bibinfo {author} {\bibfnamefont {H.}~\bibnamefont {Bentmann}}, \bibinfo {author} {\bibfnamefont {S.}~\bibnamefont {Abdelouahed}}, \bibinfo {author} {\bibfnamefont {M.}~\bibnamefont {Mulazzi}}, \bibinfo {author} {\bibfnamefont {J.~J.}\ \bibnamefont {Henk}},\ and\ \bibinfo {author} {\bibfnamefont {F.}~\bibnamefont {Reinert}},\ }\href {https://doi.org/10.1103/PhysRevLett.108.196801} {\bibfield  {journal} {\bibinfo  {journal} {Physical Review Letters}\ }\textbf {\bibinfo {volume} {108}},\ \bibinfo {pages} {196801} (\bibinfo {year} {2012})}\BibitemShut {NoStop}%
\bibitem [{\citenamefont {Zhen}\ \emph {et~al.}(2015)\citenamefont {Zhen}, \citenamefont {Hsu}, \citenamefont {Igarashi}, \citenamefont {Lu}, \citenamefont {Kaminer}, \citenamefont {Pick}, \citenamefont {Chua}, \citenamefont {Joannopoulos},\ and\ \citenamefont {Soljačić}}]{Zhen2015}%
  \BibitemOpen
  \bibfield  {author} {\bibinfo {author} {\bibfnamefont {B.}~\bibnamefont {Zhen}}, \bibinfo {author} {\bibfnamefont {C.~W.}\ \bibnamefont {Hsu}}, \bibinfo {author} {\bibfnamefont {Y.}~\bibnamefont {Igarashi}}, \bibinfo {author} {\bibfnamefont {L.}~\bibnamefont {Lu}}, \bibinfo {author} {\bibfnamefont {I.}~\bibnamefont {Kaminer}}, \bibinfo {author} {\bibfnamefont {A.}~\bibnamefont {Pick}}, \bibinfo {author} {\bibfnamefont {S.-L.}\ \bibnamefont {Chua}}, \bibinfo {author} {\bibfnamefont {J.~D.}\ \bibnamefont {Joannopoulos}},\ and\ \bibinfo {author} {\bibfnamefont {M.}~\bibnamefont {Soljačić}},\ }\href {https://doi.org/10.1038/nature14889} {\bibfield  {journal} {\bibinfo  {journal} {Nature}\ }\textbf {\bibinfo {volume} {525}},\ \bibinfo {pages} {354} (\bibinfo {year} {2015})}\BibitemShut {NoStop}%
\bibitem [{\citenamefont {Eleuch}\ and\ \citenamefont {Rotter}(2017)}]{Eleuch2017}%
  \BibitemOpen
  \bibfield  {author} {\bibinfo {author} {\bibfnamefont {H.}~\bibnamefont {Eleuch}}\ and\ \bibinfo {author} {\bibfnamefont {I.}~\bibnamefont {Rotter}},\ }\href {https://doi.org/10.1103/PhysRevA.95.022117} {\bibfield  {journal} {\bibinfo  {journal} {Physical Review A}\ }\textbf {\bibinfo {volume} {95}},\ \bibinfo {pages} {022117} (\bibinfo {year} {2017})}\BibitemShut {NoStop}%
\bibitem [{\citenamefont {Kozii}\ and\ \citenamefont {Fu}(2024)}]{Kozii2024}%
  \BibitemOpen
  \bibfield  {author} {\bibinfo {author} {\bibfnamefont {V.}~\bibnamefont {Kozii}}\ and\ \bibinfo {author} {\bibfnamefont {L.}~\bibnamefont {Fu}},\ }\href {https://doi.org/10.1103/PhysRevB.109.235139} {\bibfield  {journal} {\bibinfo  {journal} {Physical Review B}\ }\textbf {\bibinfo {volume} {109}},\ \bibinfo {pages} {235139} (\bibinfo {year} {2024})}\BibitemShut {NoStop}%
\bibitem [{\citenamefont {Jo}\ \emph {et~al.}(2021)\citenamefont {Jo}, \citenamefont {Wu}, \citenamefont {Trevisan}, \citenamefont {Wang}, \citenamefont {Lee}, \citenamefont {Kuthanazhi}, \citenamefont {Schrunk}, \citenamefont {Bud’ko}, \citenamefont {Canfield}, \citenamefont {Orth},\ and\ \citenamefont {Kaminski}}]{Jo2021}%
  \BibitemOpen
  \bibfield  {author} {\bibinfo {author} {\bibfnamefont {N.~H.}\ \bibnamefont {Jo}}, \bibinfo {author} {\bibfnamefont {Y.}~\bibnamefont {Wu}}, \bibinfo {author} {\bibfnamefont {T.~V.}\ \bibnamefont {Trevisan}}, \bibinfo {author} {\bibfnamefont {L.-L.}\ \bibnamefont {Wang}}, \bibinfo {author} {\bibfnamefont {K.}~\bibnamefont {Lee}}, \bibinfo {author} {\bibfnamefont {B.}~\bibnamefont {Kuthanazhi}}, \bibinfo {author} {\bibfnamefont {B.}~\bibnamefont {Schrunk}}, \bibinfo {author} {\bibfnamefont {S.~L.}\ \bibnamefont {Bud’ko}}, \bibinfo {author} {\bibfnamefont {P.~C.}\ \bibnamefont {Canfield}}, \bibinfo {author} {\bibfnamefont {P.~P.}\ \bibnamefont {Orth}},\ and\ \bibinfo {author} {\bibfnamefont {A.}~\bibnamefont {Kaminski}},\ }\href {https://doi.org/10.1038/s41467-021-27277-6} {\bibfield  {journal} {\bibinfo  {journal} {Nature Communications}\ }\textbf {\bibinfo {volume} {12}},\ \bibinfo {pages} {7169} (\bibinfo {year} {2021})}\BibitemShut {NoStop}%
\bibitem [{\citenamefont {Hahn}\ \emph {et~al.}(2021)\citenamefont {Hahn}, \citenamefont {Sohn}, \citenamefont {Kim}, \citenamefont {Kim}, \citenamefont {Huh}, \citenamefont {Kim}, \citenamefont {Kyung}, \citenamefont {Kim}, \citenamefont {Kim}, \citenamefont {Kim}, \citenamefont {Noh}, \citenamefont {Shim},\ and\ \citenamefont {Kim}}]{Hahn2021}%
  \BibitemOpen
  \bibfield  {author} {\bibinfo {author} {\bibfnamefont {S.}~\bibnamefont {Hahn}}, \bibinfo {author} {\bibfnamefont {B.}~\bibnamefont {Sohn}}, \bibinfo {author} {\bibfnamefont {M.}~\bibnamefont {Kim}}, \bibinfo {author} {\bibfnamefont {J.~R.}\ \bibnamefont {Kim}}, \bibinfo {author} {\bibfnamefont {S.}~\bibnamefont {Huh}}, \bibinfo {author} {\bibfnamefont {Y.}~\bibnamefont {Kim}}, \bibinfo {author} {\bibfnamefont {W.}~\bibnamefont {Kyung}}, \bibinfo {author} {\bibfnamefont {M.}~\bibnamefont {Kim}}, \bibinfo {author} {\bibfnamefont {D.}~\bibnamefont {Kim}}, \bibinfo {author} {\bibfnamefont {Y.}~\bibnamefont {Kim}}, \bibinfo {author} {\bibfnamefont {T.~W.}\ \bibnamefont {Noh}}, \bibinfo {author} {\bibfnamefont {J.~H.}\ \bibnamefont {Shim}},\ and\ \bibinfo {author} {\bibfnamefont {C.}~\bibnamefont {Kim}},\ }\href {https://doi.org/10.1103/PhysRevLett.127.256401} {\bibfield  {journal} {\bibinfo  {journal} {Physical Review Letters}\ }\textbf {\bibinfo {volume} {127}},\ \bibinfo {pages} {256401} (\bibinfo {year}
  {2021})}\BibitemShut {NoStop}%
\bibitem [{\citenamefont {Bergholtz}\ \emph {et~al.}(2021)\citenamefont {Bergholtz}, \citenamefont {Budich},\ and\ \citenamefont {Kunst}}]{Bergholtz2021}%
  \BibitemOpen
  \bibfield  {author} {\bibinfo {author} {\bibfnamefont {E.~J.}\ \bibnamefont {Bergholtz}}, \bibinfo {author} {\bibfnamefont {J.~C.}\ \bibnamefont {Budich}},\ and\ \bibinfo {author} {\bibfnamefont {F.~K.}\ \bibnamefont {Kunst}},\ }\href {https://doi.org/10.1103/RevModPhys.93.015005} {\bibfield  {journal} {\bibinfo  {journal} {Reviews of Modern Physics}\ }\textbf {\bibinfo {volume} {93}},\ \bibinfo {pages} {015005} (\bibinfo {year} {2021})}\BibitemShut {NoStop}%
\bibitem [{\citenamefont {Cayao}(2023)}]{Cayao2023}%
  \BibitemOpen
  \bibfield  {author} {\bibinfo {author} {\bibfnamefont {J.}~\bibnamefont {Cayao}},\ }\href {https://doi.org/10.1088/1361-648X/acc7e9} {\bibfield  {journal} {\bibinfo  {journal} {Journal of Physics: Condensed Matter}\ }\textbf {\bibinfo {volume} {35}},\ \bibinfo {pages} {254002} (\bibinfo {year} {2023})}\BibitemShut {NoStop}%
\bibitem [{\citenamefont {Yoshida}\ \emph {et~al.}(2018)\citenamefont {Yoshida}, \citenamefont {Peters},\ and\ \citenamefont {Kawakami}}]{Yoshida2018}%
  \BibitemOpen
  \bibfield  {author} {\bibinfo {author} {\bibfnamefont {T.}~\bibnamefont {Yoshida}}, \bibinfo {author} {\bibfnamefont {R.}~\bibnamefont {Peters}},\ and\ \bibinfo {author} {\bibfnamefont {N.}~\bibnamefont {Kawakami}},\ }\href {https://doi.org/10.1103/PhysRevB.98.035141} {\bibfield  {journal} {\bibinfo  {journal} {Physical Review B}\ }\textbf {\bibinfo {volume} {98}},\ \bibinfo {pages} {035141} (\bibinfo {year} {2018})}\BibitemShut {NoStop}%
\bibitem [{\citenamefont {Kawabata}\ \emph {et~al.}(2019)\citenamefont {Kawabata}, \citenamefont {Shiozaki}, \citenamefont {Ueda},\ and\ \citenamefont {Sato}}]{Kawabata2019}%
  \BibitemOpen
  \bibfield  {author} {\bibinfo {author} {\bibfnamefont {K.}~\bibnamefont {Kawabata}}, \bibinfo {author} {\bibfnamefont {K.}~\bibnamefont {Shiozaki}}, \bibinfo {author} {\bibfnamefont {M.}~\bibnamefont {Ueda}},\ and\ \bibinfo {author} {\bibfnamefont {M.}~\bibnamefont {Sato}},\ }\href {https://doi.org/10.1103/PhysRevX.9.041015} {\bibfield  {journal} {\bibinfo  {journal} {Physical Review X}\ }\textbf {\bibinfo {volume} {9}},\ \bibinfo {pages} {041015} (\bibinfo {year} {2019})}\BibitemShut {NoStop}%
\bibitem [{\citenamefont {Bergholtz}\ and\ \citenamefont {Budich}(2019)}]{Bergholtz2019}%
  \BibitemOpen
  \bibfield  {author} {\bibinfo {author} {\bibfnamefont {E.~J.}\ \bibnamefont {Bergholtz}}\ and\ \bibinfo {author} {\bibfnamefont {J.~C.}\ \bibnamefont {Budich}},\ }\href {https://doi.org/10.1103/PhysRevResearch.1.012003} {\bibfield  {journal} {\bibinfo  {journal} {Physical Review Research}\ }\textbf {\bibinfo {volume} {1}},\ \bibinfo {pages} {012003} (\bibinfo {year} {2019})}\BibitemShut {NoStop}%
\bibitem [{\citenamefont {Nagai}\ \emph {et~al.}(2020)\citenamefont {Nagai}, \citenamefont {Qi}, \citenamefont {Isobe}, \citenamefont {Kozii},\ and\ \citenamefont {Fu}}]{Nagai2020}%
  \BibitemOpen
  \bibfield  {author} {\bibinfo {author} {\bibfnamefont {Y.}~\bibnamefont {Nagai}}, \bibinfo {author} {\bibfnamefont {Y.}~\bibnamefont {Qi}}, \bibinfo {author} {\bibfnamefont {H.}~\bibnamefont {Isobe}}, \bibinfo {author} {\bibfnamefont {V.}~\bibnamefont {Kozii}},\ and\ \bibinfo {author} {\bibfnamefont {L.}~\bibnamefont {Fu}},\ }\href {https://doi.org/10.1103/PhysRevLett.125.227204} {\bibfield  {journal} {\bibinfo  {journal} {Physical Review Letters}\ }\textbf {\bibinfo {volume} {125}},\ \bibinfo {pages} {227204} (\bibinfo {year} {2020})}\BibitemShut {NoStop}%
\bibitem [{\citenamefont {Li}\ \emph {et~al.}(2025)\citenamefont {Li}, \citenamefont {Wang}, \citenamefont {Zhang}, \citenamefont {Liu}, \citenamefont {Liu}, \citenamefont {Chen}, \citenamefont {Deng}, \citenamefont {Ma}, \citenamefont {Polley}, \citenamefont {Thiagarajan}, \citenamefont {Kim}, \citenamefont {Yin}, \citenamefont {Shi}, \citenamefont {Xiang},\ and\ \citenamefont {Tjernberg}}]{CongLi2025}%
  \BibitemOpen
  \bibfield  {author} {\bibinfo {author} {\bibfnamefont {C.}~\bibnamefont {Li}}, \bibinfo {author} {\bibfnamefont {Y.}~\bibnamefont {Wang}}, \bibinfo {author} {\bibfnamefont {J.}~\bibnamefont {Zhang}}, \bibinfo {author} {\bibfnamefont {G.}~\bibnamefont {Liu}}, \bibinfo {author} {\bibfnamefont {H.}~\bibnamefont {Liu}}, \bibinfo {author} {\bibfnamefont {W.}~\bibnamefont {Chen}}, \bibinfo {author} {\bibfnamefont {H.}~\bibnamefont {Deng}}, \bibinfo {author} {\bibfnamefont {W.}~\bibnamefont {Ma}}, \bibinfo {author} {\bibfnamefont {C.}~\bibnamefont {Polley}}, \bibinfo {author} {\bibfnamefont {B.}~\bibnamefont {Thiagarajan}}, \bibinfo {author} {\bibfnamefont {T.~K.}\ \bibnamefont {Kim}}, \bibinfo {author} {\bibfnamefont {J.}~\bibnamefont {Yin}}, \bibinfo {author} {\bibfnamefont {Y.}~\bibnamefont {Shi}}, \bibinfo {author} {\bibfnamefont {T.}~\bibnamefont {Xiang}},\ and\ \bibinfo {author} {\bibfnamefont {O.}~\bibnamefont {Tjernberg}},\ }\href {https://doi.org/10.1002/adma.202419559} {\bibfield  {journal} {\bibinfo
  {journal} {Advanced Materials}\ }\textbf {\bibinfo {volume} {37}},\ \bibinfo {pages} {2419559} (\bibinfo {year} {2025})}\BibitemShut {NoStop}%
\end{thebibliography}

%

\end{document}